\documentclass[acmsmall]{acmart}

\AtBeginDocument{%
  \providecommand\BibTeX{{%
    Bib\TeX}}}

\setcopyright{acmlicensed}
\copyrightyear{2025}
\acmYear{2025}
\acmDOI{XXXXXXX.XXXXXXX}
\acmConference[Conference acronym 'XX]{Make sure to enter the correct
  conference title from your rights confirmation email}{June 03--05,
  2018}{Woodstock, NY}
\acmISBN{978-1-4503-XXXX-X/2018/06}

\PassOptionsToPackage{numbers,sort&compress}{natbib}
\usepackage[numbers,sort&compress]{natbib} 

\usepackage{amsmath,amssymb,amsfonts}
\usepackage{algorithmic}
\PassOptionsToPackage{off}{epstopdf}
\usepackage{graphicx}
\usepackage{textcomp}

\usepackage{xcolor}

\usepackage{threeparttable}
\def\BibTeX{{\rm B\kern-.05em{\sc i\kern-.025em b}\kern-.08em
    T\kern-.1667em\lower.7ex\hbox{E}\kern-.125emX}}
    
\usepackage[skins]{tcolorbox}
\usepackage{multirow}

\usepackage{url}  
\usepackage{listings}    
\usepackage{balance} 
\usepackage{color}

\usepackage{booktabs}
\usepackage{array}
\usepackage{tabularx}
\usepackage{siunitx}
\usepackage{makecell}
\usepackage{colortbl}
\usepackage{rotating}
\usepackage{caption}
\usepackage{tcolorbox}
\tcbuselibrary{breakable}
\usepackage{subcaption}

\newcolumntype{L}{>{\raggedright\arraybackslash}p{4.1cm}}
\newcolumntype{C}{>{\centering\arraybackslash}p{0.42cm}}

\definecolor{dkgreen}{rgb}{0,0.6,0}
\definecolor{gray}{rgb}{0.5,0.5,0.5}
\definecolor{mauve}{rgb}{0.58,0,0.82}

\tcbset{
  my box/.style={
    enhanced,
    colframe=#1!80,
    colback=#1!10,
    attach boxed title to top left={xshift=0.2cm, yshift=-0.2cm},
    boxed title style={
      colback=#1!80,
      outer arc=0pt,
      arc=0pt,
      top=0pt,
      bottom=0pt,
    },
  },
}

\newtcolorbox{summarybox}{
    enhanced, 
    colback=lightgray,
    colframe=lightgray,
    boxrule=1.2pt,
    left=1em,
    right=1em,
    top=0.5em,
    bottom=0.5em,
    borderline west={3pt}{0pt}{darkgray}, 
    sharp corners,
    breakable
}

\newtcolorbox{result-rq}[1]{
  my box=black,
  title=#1,
  boxrule=1.2pt,top=6pt,bottom=3.5pt,left=6pt,right=6pt
}

\usepackage{fontawesome5}  

\tcbset{
  myfinding/.style={
    enhanced,
    breakable,
    colback=white,
    colframe=black,
    boxrule=0.6pt,
    arc=6pt,             
    left=6pt,right=6pt,  
    top=4pt,bottom=4pt,
    sharp corners=all,   
  }
}

\usepackage{hyperref}  

\begin{document}

\title{RepoSummary: Feature-Oriented Summarization and Documentation Generation for Code Repositories}

\author{Yifeng Zhu}
\email{2501112121@stu.pku.edu.cn}
\orcid{0009-0009-0487-8404}
\affiliation{%
  \institution{Peking University}
  \city{Beijing}
  \country{China}
}

\author{Xianlin Zhao}
\email{zhaoxianlin@pku.edu.cn}
\affiliation{%
  \institution{Peking University}
  \city{Beijing}
  \country{China}
}

\author{Xutian Li}
\email{xtli25@stu.pku.edu.cn}
\affiliation{%
  \institution{Peking University}
  \city{Beijing}
  \country{China}
}

\author{Yanzhen Zou}
\email{zouyz@pku.edu.cn}
\affiliation{%
  \institution{Peking University}
  \city{Beijing}
  \country{China}
}

\author{Haizhuo Yuan}
\email{yuanhz@stu.pku.edu.cn}
\affiliation{%
  \institution{Peking University}
  \city{Beijing}
  \country{China}
}

\author{Yue Wang}
\email{wangyue0502@pku.edu.cn}
\affiliation{%
  \institution{Peking University}
  \city{Beijing}
  \country{China}
}

\author{Bing Xie}
\email{xiebing@pku.edu.cn}
\affiliation{%
  \institution{Peking University}
  \city{Beijing}
  \country{China}
}

\renewcommand{\shortauthors}{Yifeng Zhu, Xianlin Zhao, Xutian Li, Yanzhen Zou, Yue Wang, Bing Xie}

\begin{abstract}

Repository summarization is a crucial research question in development and maintenance for software engineering. 
Existing repository summarization techniques primarily focus on summarizing code according to the directory tree, which is insufficient for tracing high-level features to the methods that collaboratively implement them. 
To address these limitations, we propose RepoSummary, a feature-oriented code repository summarization approach that simultaneously generates repository documentation automatically. Furthermore, it establishes more accurate traceability links from functional features to the corresponding code elements, enabling developers to rapidly locate relevant methods and files during code comprehension and maintenance.
Comprehensive experiments against the state-of-the-art baseline (HGEN) demonstrate that RepoSummary achieves higher feature coverage and more accurate traceability. On average, it increases the rate of completely covered features in manual documentation from 61.2\% to 71.1\%, improves file-level traceability recall from 29.9\% to 53.0\%, and generates documentation that is more conceptually consistent, easier to understand, and better formatted than that produced by existing approaches.
\end{abstract}



\maketitle

\section{Introduction}
\label{introduction}
With the continuous expansion of software systems, the growing functional complexity and increasing module coupling in software repositories have made code comprehension and maintenance progressively more challenging ~\cite{jss/WoodfieldSD81, icse/XiaBLXHL18}. 
In this context, developers often devote considerable effort to understanding repositories, where high-quality documentation plays a critical role in reducing the time required for code comprehension ~\cite{sigdoc/SouzaAO05}.
However, it is time-consuming and labor-intensive for  developers to write software documentation manually ~\cite{jss/ZhiGSGSR15,tse/McBurneyJKKAMM18}. 
As a result, how to automatically summarize code at the repository-level and generate software documentation has become an important research question in the field of software engineering.

\begin{figure*}
    \centering
    \includegraphics[width=1\linewidth]{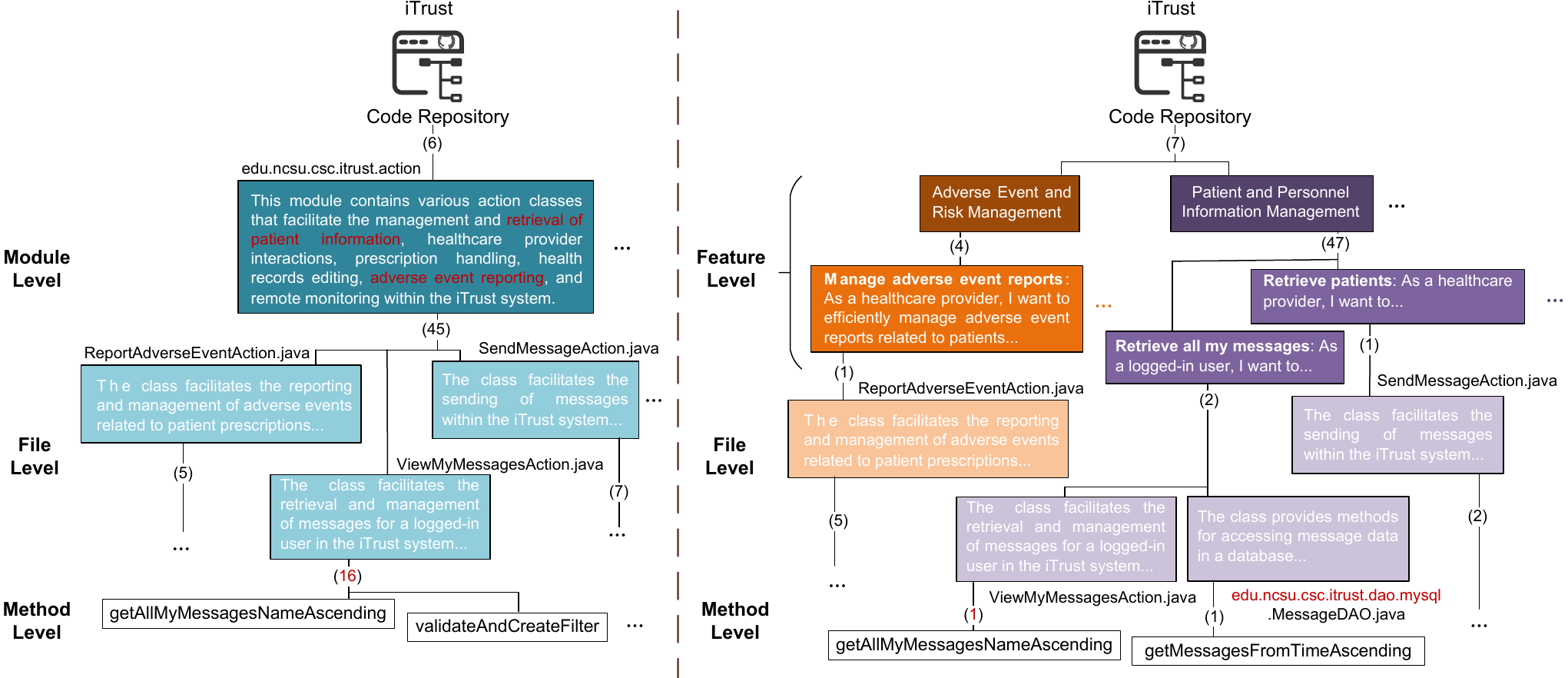}
    \caption{Differences between traditional and feature-oriented repository summarization.}
    \label{fig:compare}
\end{figure*}

Considering the complexity of repository code, researchers have leveraged Large Language Models (LLMs) to enable repository-level code comprehension, opening up new frontiers in software development and maintenance. 
These research works can generate documentation for entire repositories with LLMs through carefully designed prompts and context retrieval strategies ~\cite{ma2024alibaba, luo2024repoagent, dhulshette2025hierarchical, yang2025docagent, dearstyne2024supporting}. 
Typically, RepoAgent ~\cite{luo2024repoagent} can automatically generate functionality descriptions, parameter descriptions, and usage examples for code methods.
HMCS ~\cite{sun2025commenting} can generate summaries for each file code and then produce a module summary based on these file code summaries.
HGEN ~\cite{dearstyne2024supporting}, one of the state-of-the-art approaches, can not only generate hierarchical documentation, but also generate traceability links that connect it to repository code.

Although these research works have achieved promising performance, they still exhibit some notable limitations, particularly for new developers unfamiliar with software projects.
When a project’s file structure is disorganized and its modules are bloated, existing repository summarization approaches perform poorly and fail to help developers understand program functionalities, modules, and overall architecture. As shown in Fig.~\ref{fig:compare}, the module \texttt{edu.ncsu.csc.itrust.action} in  \texttt{iTrust} repository contains 45 files, approximately one third of the files in the repository. 
The summary generated by HMCS ~\cite{sun2025commenting} of the \texttt{action} module degenerates into a near-enumeration of project-wide functionalities, overlooking many functional details and obscuring the mapping between files and functional features. 
Consequently, its generated repository summarization \textbf{lacks a well-structured architectural and functional perspective}.

It is essential for developers to quickly grasp the overall architecture and core functionalities from the perspective of functional features, rather than reading through individual code files one by one. This approach aligns better with the actual workflow and mindset of developers ~\cite{dearstyne2024supporting, robillard2017demand}. Besides, in practice, a file may contain a lot of methods implementing  different functional features. A single functional feature often spans multiple methods across several files ~\cite{gamma1995design}.
In order to help developers locate related methods or files rapidly according to their development requirements, there is a \textbf{need to build accurate mapping relationships between functional features and code methods} in order to reduce the onboarding cost for new developers ~\cite{10.5555/2337223.2337254}.

To address the preceding challenges, we propose \textbf{RepoSummary}, a feature-oriented code repository summarization approach that simultaneously generates repository documentation automatically. 
For a large-scale code repository, we first generate two adjacency matrices based on the code dependency relationships at the file-level and the method-level.
Secondly, we summarize each method using its code and dependencies, and summarize each file using all its method summaries. Based on this, we calculate two semantic similarity matrices at the file-level and the method-level. 
Thirdly, we perform file-level clustering and method-level clustering based on the adjacency matrices and semantic similarity matrices. Each cluster of methods is a candidate functional feature.
Finally, we employ LLM to generate functional descriptions for each cluster, while simultaneously preserving the traceability links between functional features and multiple methods that collaboratively implement them throughout this process.
All the results are used to generate a systematic repository documentation.

To evaluate our approach, we conduct experiments on 3 repositories and 26 commits selected from a real development environment. These repositories are all Java projects, which contain manually-written features and traceability links between code and these features. 
Our experimental results indicate that: 
(1) Compared with the SOTA approach, RepoSummary increases the complete coverage rate of features in manual documentation from 61.2\% to 71.1\% on average. 
(2) RepoSummary enables code-functionality traceability. Compared to the baseline approach, the recall of the file traceability link improves from 29.9\% to 53.0\%.
(3) Compared with other approaches, RepoSummary is able to generate easy-to-understand, concept-consistent, and well-formatted software documentation.

To summarise, this paper makes the following main contributions:
\begin{itemize}
\item We propose RepoSummary, a feature-oriented code repository summarization approach that simultaneously generates repository documentation automatically.

\item Our approach establishes more accurate traceability links from functional features to the corresponding code elements, enabling developers to rapidly locate relevant methods and files during code comprehension and maintenance.

\item We conduct a comprehensive evaluation on open-source datasets. The results demonstrate that the documentation generated by our approach captures a larger number of functional features compared to manual documentation and provides more precise traceability links.

\end{itemize} 

The rest of the paper is organized as follows.
Section~\ref{sec:relatedwork} introduces the related work.
Section~\ref{sec:approach} introduces our proposed approach.
Section~\ref{sec:experimental setup} explains our experimental setup. Section~\ref{sec:experimental result} presents the experimental results.
Section~\ref{sec:discussion} shows the discussion. 
Finally, Section~\ref{sec:conclusion} concludes this paper.

\section{Related Work}
\label{sec:relatedwork}
In this section, we describe the related work, including (1) Code summarization and repository summarization. (2) Automated repository documentation generation.

\subsection{Code Summarization and Repository Summarization}


Automated code summarization has become an important research direction in the field of software engineering.
The main methodologies have evolved from early template-based approaches ~\cite{sridhara2010towards, sridhara2011automatically}, information retrieval (IR)-based approaches ~\cite{wong2015clocom, panichella2012mining}, to deep learning (DL)-based approaches ~\cite{iyer2016summarizing, hu2018deep, ahmad2020transformer, 9284039, 10.1145/3324884.3416578, 9678882, 9678724, leclair2019neural, lu2024improving, wan2018improving, feng-etal-2020-codebert, 10.1145/3361242.3362774, bansal2023function, hu2022practitioners}. These approaches primarily focus on method-level code ~\cite{hu2022practitioners}, generating fine-grained, low-level explanations for code methods.


With the advent of LLMs, automatic code summarization has achieved further breakthroughs. While some researchers explored using LLMs to generate method-level code summarization ~\cite{10.1145/3551349.3559548, emnlp/LomshakovPSBLN24, 10.1145/3597503.3608134, 2024arXiv240707959S}, more and more researchers are increasingly exploring LLM applications on higher-level code units, such as module-level code summarization and repository-level code summarization.
Specifically, Sun et al. ~\cite{sun2025commenting} conduct a study on how to use LLMs for commenting on higher-level code units. For module-level code, they explore generating summaries hierarchically according to the directory structure, first generating summaries for lower-level code units, and then using those summaries to generate summaries for higher-level code units.
Luo et al. ~\cite{luo2024repoagent} decompose Python code repositories into function-level units, integrating code, folder information, and code reference information into prompts to generate repository-level documentation. The documentation describes the functionality of the code, parameter details, and usage examples.
Ma et al. ~\cite{ma2024alibaba} employ a top-down approach to condense critical repository information into a knowledge graph, enabling agents to holistically understand entire repositories for downstream tasks.
Dhulshette et al. ~\cite{dhulshette2025hierarchical} propose a hierarchical summarization pipeline. They aggregate function and variable summaries into file-level descriptions, ultimately generating package-level summaries. 

Although the above works have achieved certain effectiveness, they mainly focus on the detailed descriptions of code functionality. They do not understand the project from an overall functional perspective, and their generated results are insufficient for new developers to understand the overall functionality of the project, which they are not familiar with.


\subsection{Automated Repository Documentation Generation}


In the current software engineering context, documentation is essential for understanding and maintaining repository code. To alleviate the burden of writing documentation manually, researchers have proposed various approaches to automatically generate software documentation ~\cite{nybom2018systematic}.
Many researchers focus on API documentation generation.
Specifically, Treude et al. ~\cite{treude2016augmenting} present an approach to augment API documentation with `insight sentences' from Stack Overflow that are related to a particular API type and provide insight not contained in the original API documentation.
Uddin et al. ~\cite{uddin2021automatic} present two algorithms to automatically produce API documentation from Stack Overflow by combining code examples and reviews of those examples.
Wang et al. ~\cite{wang2023gdoc} propose gDoc with the goal of handling unpopular APIs and generating structured API documentation.
Besides API documentation, Naimi et al. ~\cite{naimi2024automating} introduce a method to extract use cases from UML Use Case Diagrams and employ an LLM to generate descriptive text for extracted use cases. These use case descriptions can also be seen as a kind of software documentation.

The above research focuses on the use of APIs or the description of use cases, with a greater emphasis on the details of the code. With the development of LLMs and the expansion of their context windows, some researchers use LLMs to generate repository-level documentation.
Specifically, Dearstyne et al. ~\cite{dearstyne2024supporting} propose a multi-stage LLM framework HGEN to generate repository documentation. This framework can produce user stories in documentation and generate trace links at the file-level, therefore, readers can have a comprehensive understanding of the requirements of the entire repository. 
As mentioned earlier, Luo et al. ~\cite{luo2024repoagent} propose RepoAgent for repository-level documentation generation. Developers can understand how to leverage the code by reading the code functionality, parameter description, and usage considerations in the documentation.

It can be seen that the existing work mainly focuses on the aggregation and generation of relevant information for the use of methods or APIs, focusing on the richness and accuracy of information, which is not friendly to developers who are not familiar with the code to get a macro-level understanding of the repository.
In addition, the documentation structure is not very well-organized, and there may be duplication and redundancy among some different subsections.

\subsection{Remaining Gaps and Challenges}

To sum up, we need to put forward a functional feature-oriented repository summarization approach for new developers. The main difficulties and technical challenges of this approach are as follows:

\textbf{(1) Extracting hierarchical functional features from repository code.} A single functional feature of a code repository often spreads across multiple files, classes, and methods. These files and methods implement a functional feature collaboratively. It is important to recognize that the directory structure of a code repository does not necessarily correspond directly to its functional organization. 
Thus, in order to provide developers with a comprehensive understanding of the functionality and even allow them to reuse specific functional features, it is important to extract functional features from entire repositories. 
However, due to the context length limitation of LLMs, it is impossible to input all the code into an LLM at once. Therefore, it is a challenge to construct an appropriate repository code context in order to enable LLMs to understand each part of the code and generate comprehensive summaries.

\textbf{(2) Establishing method-level traceability between features and code.} The goal of automatic code repository summarization is to help users better understand the repository, enabling rapid code localization during the iterative development of requirements.
Although several studies have explored traceability links between software requirements and code, most of them focus primarily on file-level traceability. However, they provide limited support for finer-grained method-level traceability, and the accuracy of such links still leaves substantial room for improvement.

\begin{figure}
    \centering
    \includegraphics[width=0.98\linewidth]{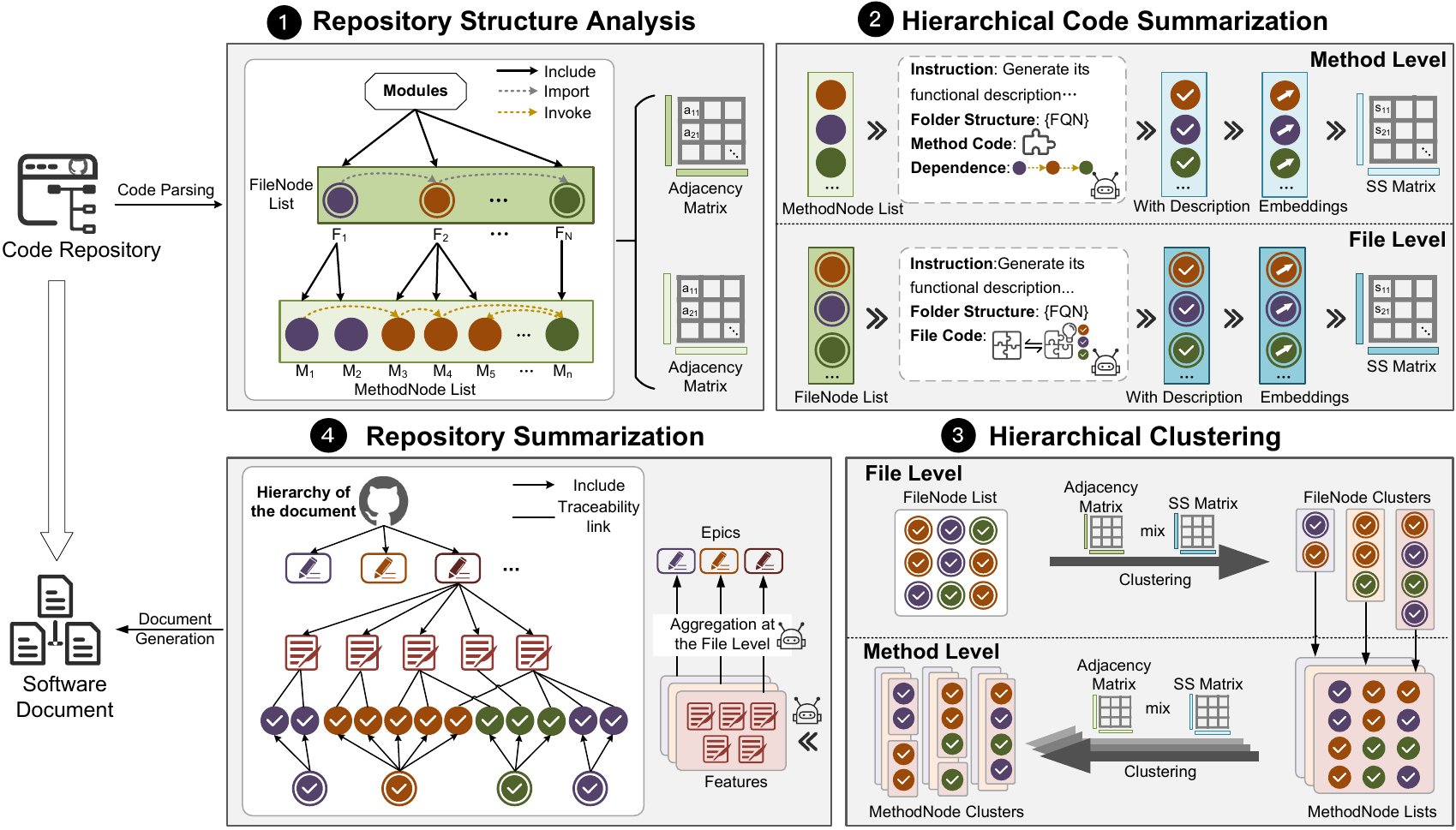}
    \caption{The framework of RepoSummary.}
    \label{fig:overview}
\end{figure}

\section{Approach}
In this section, we describe the design of RepoSummary for hierarchical feature-oriented summarization and documentation generation of code repositories. As illustrated in Fig.~\ref{fig:overview}, the approach consists of the following four phases: 
\begin{itemize}
\item{\textbf{Repository structure analysis}}: For a large-scale code repository, we first extract the code dependency relationship at the file-level and method-level, and generate different-level adjacency matrices.

\item{\textbf{Hierarchical code summarization}}: We summarize each method using its code and dependencies, and summarize each file using all its method summaries. Based on this, calculate a file-level semantic similarity matrix and a method-level similarity matrix. 

\item{\textbf{Hierarchical clustering}}: We use multi-level clustering technology based on the adjacency matrix and the semantic similarity matrix. Here, file-level clustering is used to help get a more accurate method-level clustering.

\item{\textbf{Repository summarization}}: Here, we use LLM to generate features using method-level clusters, combining these features to generate epics, while preserving the traceability links between features and methods during this process. All the results are to generate a hierarchical software documentation.
\end{itemize}

\label{sec:approach}

\subsection{Repository Structure Analysis}
In the initial phase, with the aim of understanding the structure of a large-scale code repository, RepoSummary begins by parsing the repository code and extracting key information using the static analysis tool Eclipse JDT Core\footnote{https://projects.eclipse.org/projects/eclipse.jdt}. 

The key information includes entities, primarily including file nodes and method nodes, and relationships that contain mutual imports between file nodes, file nodes including method nodes, and mutual invokes between method nodes. After getting these relationships, we can generate the corresponding \textbf{adjacency matrix}. The matrix of the file layer is a symmetric matrix, where $a_{ij}$ indicates whether there is an import and imported relationship between the $i$-th file node and the $j$-th file node. If there exists such a relationship, the value is 1. And the adjacency matrix of the method layer is generated the same way as above. The different levels of adjacency matrices are utilized throughout hierarchical clustering (step 3).

\subsection{Hierarchical Code Summarization}

To summarize the code at the repository level, we should first summarize the code at the method-level and then file-level. At the method-level, we follow ~\cite{luo2024repoagent} to generate summaries the original code by extracting the original code, location represented as a fully qualified name (FQN), and its related code from all interconnected method nodes. These elements are then integrated into a prompt to produce a concise summary.
To preserve method details, we partition each method summary into three components: (1) description, which outlines the method’s functionality; (2) workflow, which details the internal execution steps; and (3) quality, which specifies performance and other non-functional requirements. 
Then we employ the Sentence-BERT technology~\footnote{https://www.sbert.net/} to vectorize each method node's descriptions, producing embeddings that we use to construct a \textbf{Semantic similarity matrix (SS matrix)}. It characterizes functional similarity between methods. In this matrix, $s_{ij}$ denotes the semantic similarity between the $i$-th method node summary and the $j$-th method node summary.

 For file-level code summarization, we design a robust summarization workflow. If a file fits within the LLM’s context window, we directly include its full source code and location in the prompt to generate the file-level summary. When the file exceeds the context window, we replace each method in the source with its previously generated summary to reduce the overall input length. File summary only includes outlines of the file’s functionality.

Finally, we generate the \textbf{SS matrix} at the file node level. The generation process is similar to that at the method level. The different levels of SS matrices are also utilized throughout hierarchical clustering (step 3).

\subsection{Hierarchical Clustering}
Owing to the limited context capacity of LLM, it is necessary to partition the code repository. The aim of the partition is to aggregate code with similar functionalities or those collaboratively implementing a specific function into cohesive clusters. To achieve this goal and to also establish more accurate traceability links, we propose a hierarchical clustering technology that leverages the Leiden ~\cite{traag2019louvain} community detection algorithm to perform clustering at both the file and method levels. Leiden algorithm is well-suited for large-scale, weighted, and sparse graphs. It supports multiple objective functions to accommodate diverse optimization goals, and the accompanying Leiden library is mature and stable, facilitating sub-sampling and consensus procedures in practical development environments, thereby improving experimental reproducibility. Leiden optimizes a chosen community quality function (e.g., CPM or modularity/RB) using an iterative three-phase procedure (local moving, refinement, aggregation) that guarantees well-connected communities, improves speed, and avoids badly connected subcommunities.
For this experiment, we chose the community detection method of cluster partitioning modularity (CPM) because the scale of associated files and functions can vary significantly across different features. Standard modularity tends to generate communities of similar sizes. CPM better captures small communities by adjusting the $\gamma$ parameter.

 The algorithm clusters a graph by maximizing its modularity, which is defined as:
\begin{equation}
\label{CPM}
    Q_\gamma= \sum_{c} \left( \sum_{i<j,i,j\in c} A_{ij} - \gamma\binom{n_c}{2}\right)
\end{equation}

where, $A_{ij}$ is the weight of edge from node $i$ to $j$; $c$ indexes communities; $n_c$ is the number of nodes in community $c$; $\gamma $  is the resolution parameter.


In order to automatically obtain the optimal parameter $\gamma$, we choose stability, separation, and small-cluster fraction to form the final score:

\textbf{Stability}: The mean Adjusted Rand Index of all pairwise splits, with high ARI indicating strong robustness to randomness under the same $\gamma$.

\textbf{Separation}: This metric is computed by subtracting the within-cluster mean (weighted by cluster size) from the out-of-cluster mean, thereby quantifying the degree to which clusters are separable.

\textbf{Small-Cluster Fraction}: This metric measures the proportion of single-point clusters as well as the proportion of very small clusters (with size $\leq$ 2). A high fraction of such clusters typically indicates severe imbalance division of software files or methods, which may render the resulting structure less operable and difficult to maintain.

\begin{figure}
    \centering
    \includegraphics[width=0.98\linewidth]{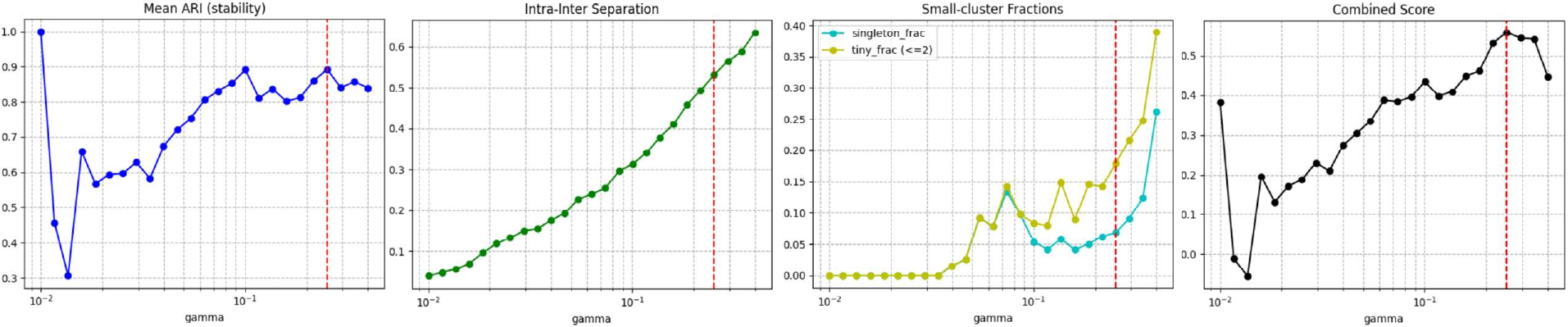}
    \caption{Stability, Separation, Small-Cluster Fraction and Combined Score in different $\gamma$. }
    \label{fig:gamma}
\end{figure}

The combined score is the stability value plus the separation decreasing cluster score.
We iterate over $\gamma$ on a logarithmic scale, resulting in a plot Fig.~\ref{fig:gamma}. The red line part indicates the best $\gamma$.

First, we cluster at the file-level. The weight matrix of clustering is obtained by mixing the adjacency matrix and the SS matrix. Then, Leiden algorithm is used for clustering, and our auto-tuning method is used to find the best $\gamma$ to obtain a series of file clusters.

Second, we expand the file clusters to the method layer and perform clustering in each expanded file cluster separately. As same as the file-level, we mix the adjacency matrix and the SS matrix. A series of method clusters is obtained after clustering.

\subsection{Repository Summarization}
\begin{figure}
    \centering
    \includegraphics[width=0.98\linewidth]{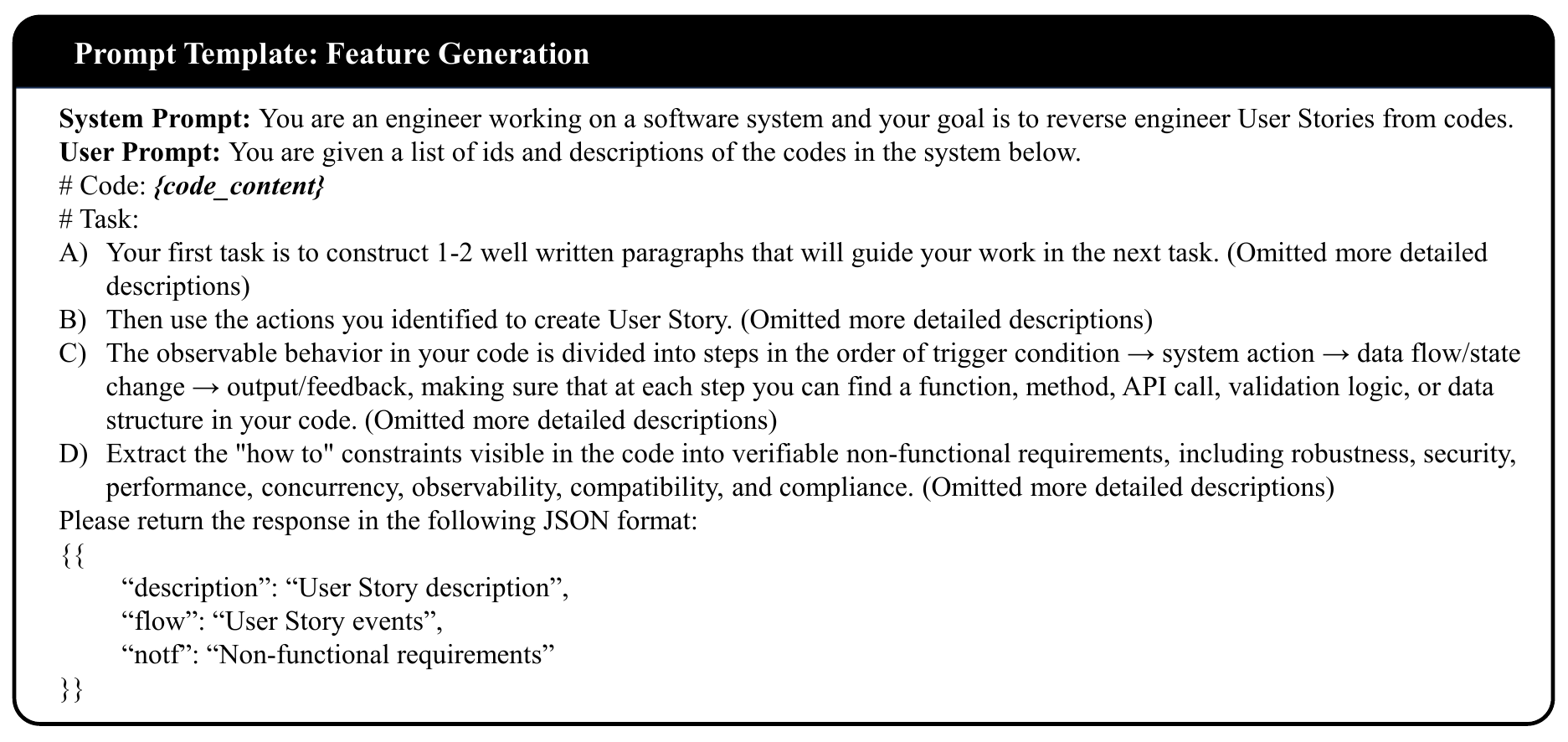}
    \caption{The prompt template for Feature generation}
    \label{fig:prompt}
\end{figure}

In this phase, we first try to summarize every method cluster to get its functional features. The prompt design is shown in Fig.~\ref{fig:prompt}.
Here, we use method summary instead of method code to represent each method in the cluster because a cluster may include too many methods and exceed the context limit of LLM.
At the same time, we adopt the Chain-of-Thought (CoT) prompting strategy to enhance the LLM’s ability to generate features. Specifically, we design a series of tasks that guide the LLM to perform step-by-step reasoning during the generation process.

Second, we aggregate features belonging to the same file cluster to form an epic. We directly populate the prompt with features belonging to the same file cluster, and then let LLM generate a common epic. For example, "Retrieve all my messages" and "Retrieve patients" are aggregated to form "Patient and Personnel information Management".

Based on the summarized functional features, epics, and their traceability link to the method code, we can finally generate a hierarchical software documentation. 
At the top level, the documentation is organized by epics. Under each epic, we present the corresponding features, followed by the associated files and methods. 
Notably, a file is linked to a feature only when the feature is relevant to at least one method in the file. 

\section{Evaluation Setup}
\label{sec:experimental setup}

RepoSummary leverages repository code to generate functional feature-oriented documentation that aligns with developer practices. To evaluate our approach, we seek to answer the following four research questions: 
\begin{enumerate}
    \item[\textbf{RQ1:}] \textbf{Feature Coverage Evaluation} - To what extent do the features extracted by RepoSummary cover and elaborate on the manual documentation?
    \item[\textbf{RQ2:}]  \textbf{Traceability Link Evaluation} - How effectively does RepoSummary establish and maintain traceability between features and the underlying code elements?

    \item[\textbf{RQ3:}] \textbf{Helpfulness Evaluation} - To what degree does the generated documentation assist developers in realistic development environments and tasks? 

    \item[\textbf{RQ4:}]  \textbf{Ablation Study} - How do different constructions of the clustering influence feature coverage and the quality of traceability links? 
\end{enumerate}

\subsection{Comparing Approaches}
To enable a comprehensive comparison, we evaluate our approach and two state-of-the-art LLM-based repository summarization approaches:

(1) \textbf{HMCS} by Sun et al. in 2025 ~\cite{sun2025commenting}, a  recently proposed repository summarization approach that focuses on generated summaries for module-level code units. Summaries of these module-level code units are highly valuable for quickly gaining a macro-level understanding of various program functionalities, modules, and architectures.

(2) \textbf{HGEN} by Dearstyne et al. ~\cite{dearstyne2024supporting} in 2024, a state-of-the-art maintenance tool that incrementally converts source code into well-structured documentation, dynamically constructs document hierarchies, and produces traceability links connecting the document structure to files.

All these approaches used LLM for code summarization. We conduct our experiments using \textbf{GPT-4o mini }(2024-07-18) ~\cite{gpt-4o} and \textbf{Claude 3}. These models differ in context length and specialization, allowing us to examine how variations among LLMs affect approaches.

\subsection{Datasets}
\begin{table}
\centering
\caption{Overview of repository dataset.}
\label{tab:datasets}
\small
\begin{tabular}{lccccccc} 
\toprule
\multicolumn{1}{c}{\multirow{2}{*}{\textbf{Repository}}} & \multicolumn{1}{c}{\multirow{2}{*}{\textbf{Domain}}} & \multicolumn{3}{c}{\textbf{Code}}& \multicolumn{3}{c}{\textbf{Document}}\\
\cmidrule(lr){3-5} \cmidrule(lr){6-8}
 & & \textbf{LOC} & \textbf{\#File} & \textbf{\#Method} & \textbf{\#Feature} & \textbf{\#TL} & \textbf{\#TL per Feature} \\
\hline
Dronlogy\cite{dearstyne2024supporting}\cite{fuchss2025lissa} & Aerospace & 35054 & 423 & 2526 & 99 & 602  &  6.08\\ 
eTour\cite{fuchss2025lissa}   & Tourism    & 24729 & 116 & 1022 & 58  & 308  & 5.31 \\
iTrust\cite{fuchss2025lissa}  & Healthcare & 14235 & 137 & 561  & 131 & 286  & 2.18 \\
\bottomrule
\end{tabular}

\begin{tablenotes}
\footnotesize
    \item TL: traceability link.
\end{tablenotes}
\end{table}
\begin{figure}
  \centering
  \begin{subfigure}[b]{0.42\textwidth}
    \centering
    \includegraphics[width=\textwidth]{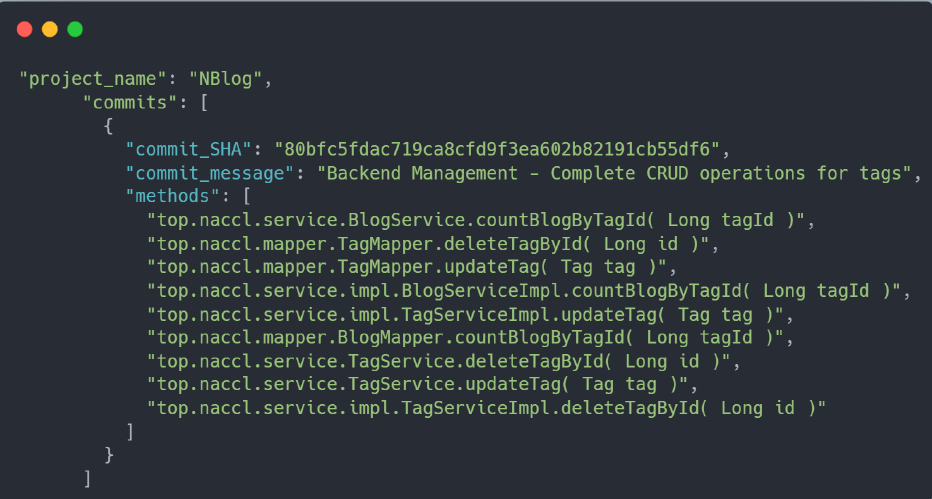}
    \caption{An example for commit dataset}
    \label{fig:commit-example}
  \end{subfigure}
  \hfill
  \begin{subfigure}[b]{0.55\textwidth}
    \centering
    \includegraphics[width=\textwidth]{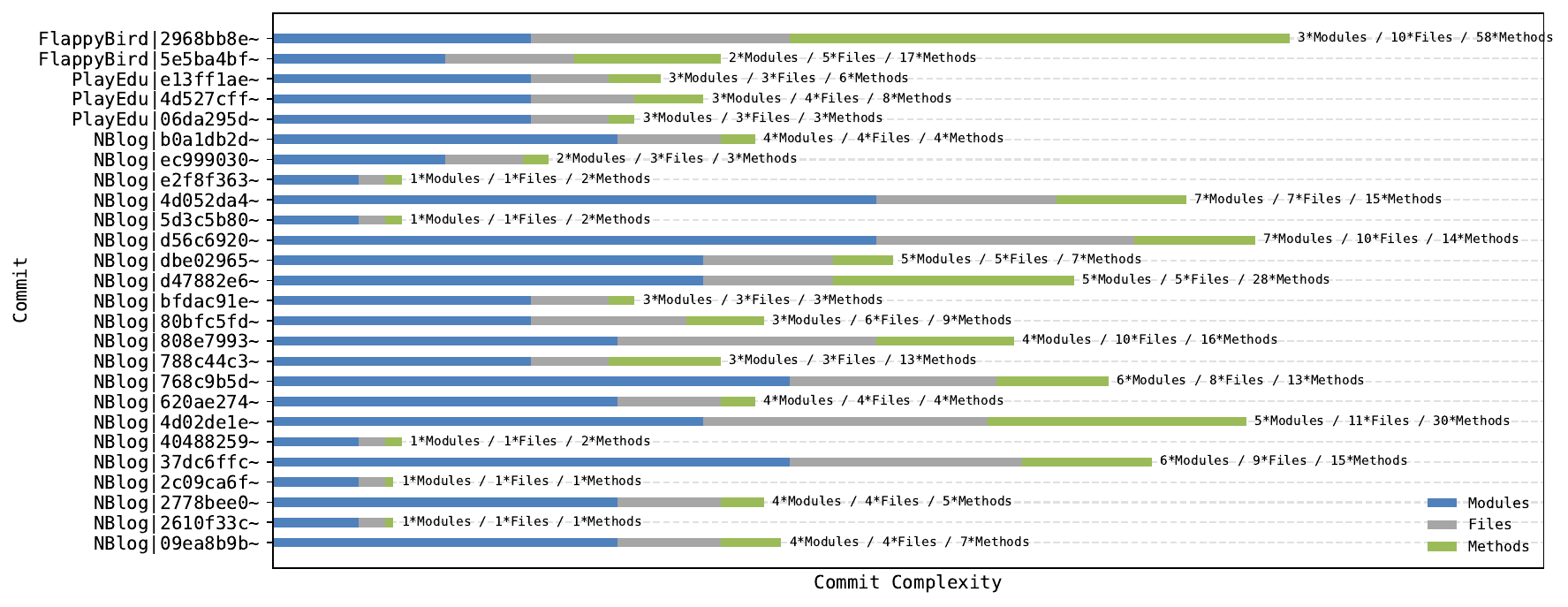}
    \caption{Commit complexity across modules, files and methods}
    \label{fig:commit-complexity}
  \end{subfigure}
  \caption{Overview of commit dataset}
  \label{fig:dataset-commit}
\end{figure}

We evaluate these approaches on 3 open-source code repositories (Repository Dataset) and 26 real-world code commit practices (Commit Dataset):

\textbf{Repository Dataset:} 
we select the repositories that can provide code artifacts and documentation that adhere to established writing standards, exhibit no obvious syntactic errors, and are widely used as benchmarks in traceability link recovery (TLR) research.
Table~\ref{tab:datasets} lists the repositories used for feature coverage and traceability-link evaluation. \texttt{Dronology} ~\cite{cleland2018dronology} is an open-source small Unmanned Aerial System (sUAS) platform for controlling and coordinating multiple sUAS in search-and-rescue, surveillance, and scientific data-collection missions. The \texttt{eTour} and \texttt{iTrust} projects, developed by the \textsl{Center of Excellence for Software \& Systems Traceability}  (CoEST)\footnote{http://sarec.nd.edu/coest/datasets.html}, target the tourism and healthcare domains, respectively. 

 \textbf{Commit Dataset:} To approximate real-world development settings, we sampled GitHub repositories from the past five years that use mature, contemporary technology stacks. We initially selected 10 projects with more than 500 stars and well-documented READMEs that guide configuration and setup. After attempting to build and run each project, we excluded 5 that could not be executed successfully, as their codebases appeared incomplete. Among the remaining 5, two contained only bug-fix commits without functional changes. Consequently, we curated 26 commits from the final three projects that introduce new features and include accompanying descriptions, ensuring that commits from different repositories do not interfere with one another. We used these commits for the helpfulness evaluation. An example of the commit data format is shown in Fig.~\ref{fig:commit-example}. Fig.~\ref{fig:commit-complexity} shows their complexity. For instance, commit \#2968bb8e~ in the \texttt{FlappyBird} repository involves 3 cross-module modifications (blue), 10 cross-file modifications (gray), and 58 cross-method modifications (green). These metrics jointly characterize the scope and interdependence of code changes, thereby reflecting the overall complexity of this commit.

\subsection{Evaluation Metrics}

\subsubsection{Feature Coverage Evaluation}

We evaluate coverage using three metrics: Covered (C), Covered by (CB), and Completely Covered (CC). 
Among them, $C$ and $CB$ are used in ~\cite{dearstyne2024supporting}, and $CC$ is used in this paper to evaluate whether the generated features can jointly cover the ground truth features.

\textbf{Covered (C)}: This metric measures the extent to which functional features in the manual documentation are covered by those in the automatically generated features. It is calculated as ~\eqref{eq:Covered}.

\begin{equation}
C = \frac{|Q_{Manual-Generated}|}{|Q_{Manual}|}
\label{eq:Covered}
\end{equation}

In this equation, $Q_{Manual}$ represents the functional features in manual documentation, and $Q_{Manual-Generated}$ represents the functional features in  manual documentation which are relevant to some of those in our automatically generated documentation. The judgement criteria of relevancy will be introduced in detail in Section~\ref{Automatic Evaluation}.

Covered includes both Partially Covered and Completely Covered.
Some ground truth features may encapsulate high-level actions (`managing various healthcare-related data'), whereas the generated features might focus on describing lower-level, more specific actions (`update the information for a hospital'). 
As a result, the generated features may convey some aspects of the meaning contained in the ground truth features, but not fully capture them.

\textbf{Completely Covered (CC)}: To assess whether the generated features can completely convey the meaning of the ground truth features, in our evaluation, we specifically compute the \textbf{$CC$} metric. The calculation is as ~\eqref{eq:CC}.

\begin{equation}
C = \frac{|Q_{Manual-C-Generated}|}{|Q_{Manual}|}
\label{eq:CC}
\end{equation}

In this equation, $Q_{Manual-C-Generated}$ represents the functional features in manual documentation whose meanings are \textbf{completely covered} by their relevant generated features. We will further give a detailed explanation in Section~\ref{Automatic Evaluation} on how to determine complete coverage.

\textbf{Covered by (CB)}: This metric denotes the proportion of the generated functional features that are covered by features in the manual documentation over all our generated features. The calculation is shown in ~\eqref{eq:CoveredBy}.

\begin{equation}
CB = \frac{|Q_{Generated-Manual}|}{|Q_{Generated}|}
\label{eq:CoveredBy}
\end{equation}

In this equation, $Q_{Generated}$ represents the functional features that our approach generates, and $Q_{Generated-Manual}$ represents the functional features in our generated documentation which are relevant to some of those in the manual documentation.

\subsubsection{Traceability Link Evaluation}

We evaluate the effectiveness of traceability links using three metrics: P, R and F1.

\textbf{Precision (P)}: Precision represents the correct proportion of automatically extracted traceability links. It is calculated as follows:

\begin{equation}
P = \frac{|TL_{Generated-Manual}|}{|TL_{Generated}|}
\label{eq:Precision}
\end{equation} 

In ~\eqref{eq:Precision}, $TL_{Generated}$ represents all the generated traceability links. $TL_{Generated-Manual}$ represents traceability links that belong to the generated features which are relevant to some of those features in manual documentation.

\textbf{Recall (R)}: Recall represents the proportion of manually extracted traceability links which are covered by automatically extracted traceability links ~\eqref{eq:Recall}:

\begin{equation}
R = \frac{|TL_{Manual-Generated}|}{|TL_{Manual}|}
\label{eq:Recall}
\end{equation} 

In ~\eqref{eq:Recall}, $TL_{Manual}$ represents all the traceability links in manual documentation. $TL_{Manual-Generated}$ represents traceability links that belong to the manual features which are relevant to some of our generated features.

\textbf{F1-Score (F1)}: F1-Score represents the harmonic mean of precision and recall and is calculated as follows:

\begin{equation}
F1 = \frac{2 * P * R}{P + R}
\label{eq:F1}
\end{equation} 

\subsection{Automatic Evaluation Based on LLM-as-a-judge}
\label{Automatic Evaluation}
\begin{figure}
    \centering
    \includegraphics[width=1\linewidth]{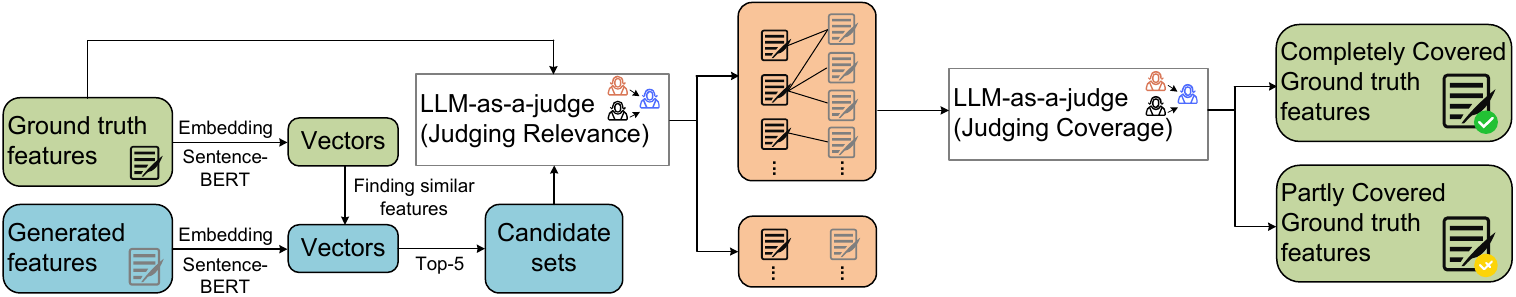}
    \caption{Overview of experiment procedures for evaluating the coverage (RQ1).}
    \label{fig:RQ}
\end{figure}

Feature coverage measures the alignment between the features generated by our approach and in manual documentation. Conventional metrics based on lexical overlap (e.g., BLEU ~\cite{bleu}) or semantic similarity (e.g., embedding-based scores ~\cite{Gao_Yao_Chen_2021}) often struggle to capture fine-grained differences in feature descriptions. Consequently, prior studies have frequently relied on costly and time-consuming user studies with manual evaluation to assess feature coverage. 
Recent work, however, has demonstrated the effectiveness of LLM-based judges for automated evaluation ~\cite{li2025generation}. For example, Lin et al. ~\cite{Lin_Chen} applied the LLM-as-a-judge paradigm to the task of article retrieval, where the judgment LLM was provided with both the conversational context and the model’s returned answer. Their approach replaces human annotation with LLM judgments, providing a practical blueprint for our setting.

Building on these insights, we propose a two-stage automatic evaluation framework to assess feature coverage in the generated documentation. All judgments follow the principles of multi-agent collaboration ~\cite{li2025generation}. 
Specifically, we employ two evaluation models (GPT-5 and Claude 4), different from the generation model serving as the initial judges. If their decisions are consistent, we adopt the result directly. In case of disagreement, we introduce DeepSeek V3.1 as a tie-breaking judge to determine the final verdict.
To enhance evaluation stability, the temperature of all judge LLMs is fixed at 0.

 \textbf{Stage I: Judging relevance.}
Developers rarely focus on a single concept when performing a development task. Therefore, we treat each feature as an integrated unit and compute feature coverage instead. 
First, we retrieve the top-5 candidate features most similar to each ground truth feature. We observed that candidates ranked sixth and below exhibit low semantic similarity to the ground truth (typically around 0.3), and excluding them also reduces the number of tokens. Prior work shows that the quality ranking of candidate responses can be artificially influenced by their ordering in the context ~\cite{Wang_Li_Chen_Zhu_Lin_Cao_Liu_Liu_Sui_2023}. Therefore, we assess relevance individually rather than jointly ~\cite{chang2023booookscore}.
Based on this, we employ an LLM-as-a-judge to assess the relevance between each ground truth feature and its candidate features. 
Our prompt design follows BOOOOKSCORE ~\cite{chang2023booookscore}, which is used to automatically evaluate summaries of long-form documents (e.g., books) using criteria that include entity omission and event omission.
The judgment includes two steps: (1) Extract the entity set and the operation set from the manual documentation (ground-truth features) and from the generated features. (2) Determine relevance by checking whether both the entity sets and the action sets have a non-empty intersection. If both intersections are non-empty, we add an association edge; otherwise, we reject the candidate.

 \textbf{Stage II: Judging Coverage.}
For each ground truth feature, we collect its related generated features and assess whether it is Completely Covered. Otherwise, it is partly covered.
As in the previous step, we first extract the entity set and the operation set. We then determine Complete Coverage by checking whether the entity and operation sets in the related generated features jointly subsume those of the ground-truth feature.

\begin{table}
\caption{Comparison of scores by human and LLM-as-a-judge for feature coverage.}
\label{tab:human-eval-coverage}
\centering
\small
\begin{tabular}{lcccccccccc}
\toprule
& \multicolumn{2}{c}{C} & \multicolumn{2}{c}{CB}& \multicolumn{2}{c}{CC} & \multirow{2}{*}{Correlation} \\
\cmidrule(lr){2-3}\cmidrule(lr){4-5}\cmidrule(lr){6-7}
 & Human & LLM  & Human & LLM  & Human & LLM    & \\
\midrule
HMCS (GPT-4o mini) & .987 & .971 & .873 & .889 & .600 & .557 & .829\\
HGEN (GPT-4o mini) & 1.00 & 1.00 & .495 & .697 & .653 & .680 & .647 \\
RepoSummary (GPT-4o mini) & 1.00 & 1.00 & .553 & .680 & .765 & .783 & .765\\
\hline

HMCS (Claude-3) & .987 & .971 & .873 & .889 & .600 & .557 & .829 \\
HGEN (Claude-3) & .971 & .962 & .425 & .700 & .665 & .687 & .771 \\
RepoSummary (Claude-3) & 1.00 & .982 & .576 & .693 & .677 & .692 & .638 \\
\midrule
Correlation &\multicolumn{2}{c}{.839} & \multicolumn{2}{c}{.590} & \multicolumn{2}{c}{.590} & .809\\
\bottomrule
\end{tabular}
\end{table}

To assess agreement between the LLM-as-a-judge and human evaluations, we report the average coverage scores from both sources in Table~\ref{tab:human-eval-coverage}. We quantify concordance using Spearman’s rank correlation coefficient ~\cite{sun2025commenting}. 
The first and second authors conducted the human evaluation, and both have more than five years of Java development experience. Because neither has prior expertise in the aerospace domain, and Dronology comprises a large number of code files that are difficult to comprehend, the human evaluation was performed only on eTour and iTrust. The evaluators thoroughly reviewed each project’s code and documentation before rating; disagreements were resolved through discussion to reach a consensus.

 As shown in Table~\ref{tab:human-eval-coverage}, the LLM-as-a-judge approach demonstrates trends consistent with human judgments, with correlation coefficients ranging from 0.590 to 0.839. Furthermore, the relative ranking of methods across different metrics remains stable, providing additional evidence for the validity of LLM-as-a-judge. 

The automatic evaluation requires approximately two hours and costs \$2.24 on average, whereas the average time for a human evaluation exceeds five hours per project. To improve efficiency, we therefore employ automatic evaluation in the subsequent experiments. 

\section{Evaluation Results}
\label{sec:experimental result}
\subsection{RQ1. Feature Coverage Evaluation } 
\begin{table}
  \caption{Feature coverage results of different approaches.}
  \label{tab:Feature coverage of different approaches}
  \centering
  \small
  \begin{tabular}{L *{12}{C}}
    \toprule
    \multirow{2}{*}{\makecell[l]{\textbf{Approach}}} &
    \multicolumn{4}{c}{Dronology} &
    \multicolumn{4}{c}{eTour} &
    \multicolumn{4}{c}{iTrust} \\
    \cmidrule(lr){2-5} \cmidrule(lr){6-9} \cmidrule(lr){10-13} 
    & \#N & C & CB & CC
    & \#N & C & CB & CC
    & \#N & C & CB & CC\\
    \midrule
    HMCS (GPT-4o mini)
     & 78 & .980 & \textbf{.756} & .535
     & 18 & \textbf{1.00} & \textbf{.944} & .526
     & 6 & .941 & \textbf{.833} & .588  \\
    HGEN (GPT-4o mini)
     & 366 & .990 & .440 & .475
     & 115 & \textbf{1.00} & .670 & \textbf{.772}
     & 65 & \textbf{1.00} & .723 & .588\\
    RepoSummary (GPT-4o mini)
     & 307 & \textbf{1.00} & .401 & \textbf{.567}
     & 51 & \textbf{1.00} & .882 & \textbf{.772}
     & 213 & \textbf{1.00} & .478 & \textbf{.794}\\
\hline
    HMCS (Claude-3)
     & 78 & .990 &\textbf{.679} & .495  
     & 18 & \textbf{1.00} & .889 & .632
     & 6 & .936 & \textbf{.833} & .324 \\
    HGEN (Claude-3)
     & 309 & \textbf{1.00} & .437  & .475
     & 106 & .982 & .670 & .756
     & 78 & .941 & .730 & .618 \\
    RepoSummary (Claude-3)
     & 368 & .990 & .359 & .495
     & 105 & \textbf{1.00} & .810 & .737
     & 59& \textbf{.964} & .576 & \textbf{.647}\\
    \bottomrule
  \end{tabular}
\end{table}
\begin{figure}
    \centering
    \includegraphics[width=0.95\linewidth]{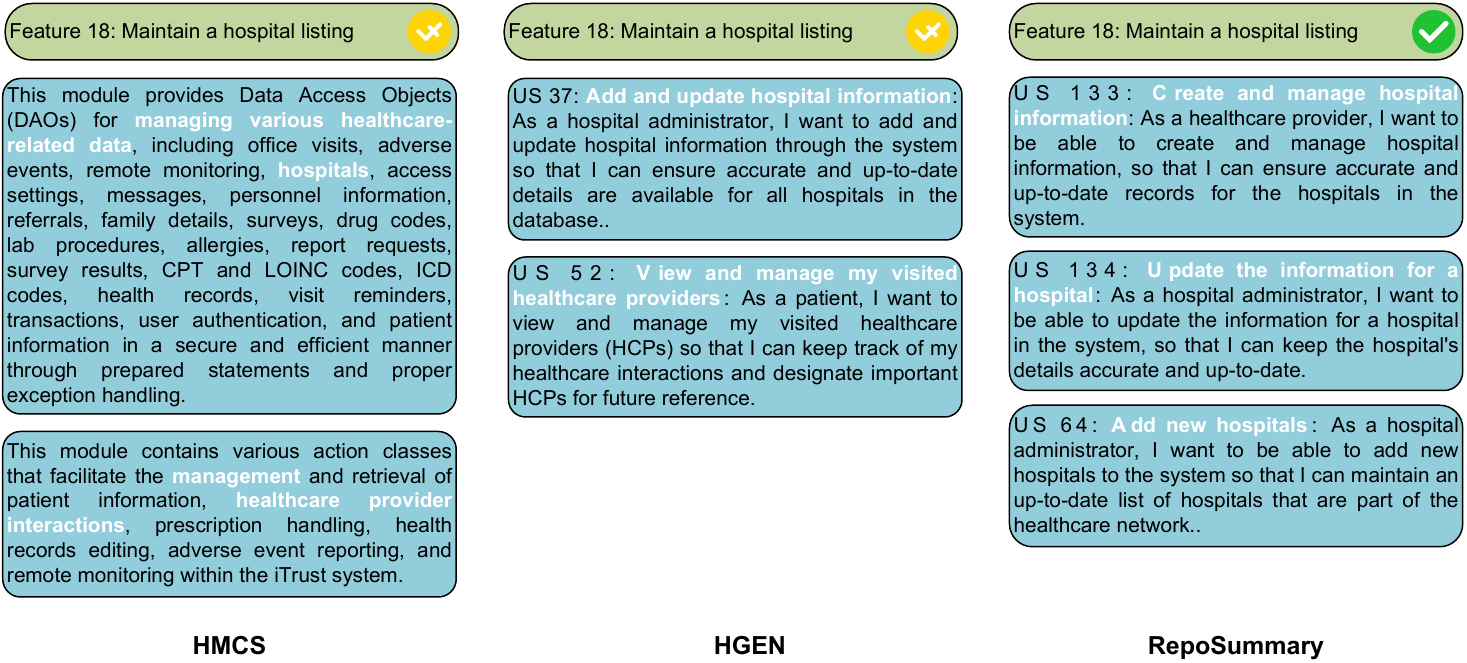}
    \caption{The difference of partly covered and completely covered.}
    \label{fig:RQ1-example}
\end{figure}
To answer RQ1, we compare the feature coverage evaluation of our approach with two baselines, HMCS and HGEN.
For all approaches, we adopt the same LLM backbone (GPT-4o mini and Claude 3) with identical decoding parameters (temperature, top-p, and max tokens), so that differences in performance arise from methodological differences rather than parameter setting.
Table \ref{tab:Feature coverage of different approaches} reports the results for three metrics: Covered (C), Covered by (CB), and Completely Covered (CC).
Seen from the table, we make the following observations:

\textbf{Improved completely covered}. In the GPT-4o mini model, our approach gains a high CC (Avg 0.711, higher than 0.612 in HGEN). In Claude, our approach gains lower CC (Avg 0.626) while HGEN gains 0.616. When a project’s architecture is well structured, HMCS, which summarizes along the directory tree, can approach the performance of our method; on \texttt{Dronology}, its CC even surpasses HGEN. 
However, in projects with entangled architectures such as \texttt{iTrust}, HMCS degrades. \texttt{iTrust} contains only six modules, yet the \texttt{action} module includes 45 of the 137 files, nearly one third of the files in the repository, causing summaries to omit many details. As illustrated in Fig.\ref{fig:RQ1-example}, \texttt{edu.ncsu.itrust.dao} effectively lists most of \texttt{iTrust}’s functionality, producing overly terse generated features that the LLM-as-a-judge rejects. HGEN, which derives features from file-level summaries, uses a finer granularity than modules, but methods in a single file may serve for multiple features; restricting summarization to the file level still overlooks important details and fails to cover all concepts in the ground truth features.

\textbf{Generated feature granularity aligns more closely with manual documentation}. In our automatic evaluation, GPT-5 and Claude 4 jointly assess the relevance and completeness between ground-truth and generated features. When their judgments diverge, we invoke DeepSeek V3.1 to review their answers and rationales and issue a final decision. We quantify inter-rater agreement between GPT-5 and Claude 4 using Cohen’s kappa. For HMCS, kappa is typically around 0.07, with a 95\% confidence interval that includes negative values. Reviewing the LLM-provided answers and rationales, we find that, relative to GPT-5, Claude 4 acts as a stricter judge. As shown in Fig.~\ref{fig:RQ1-example}, for the 18th ground-truth feature under HMCS, GPT-5 labeled CC as “yes” whereas, Claude 4 required the entity set and operation set in the generated features not only to include those in the ground truth but also to be approximately equal, resulting in a disagreement. The kappa for RepoSummary is about 0.32, and for HGEN it is about 0.28, indicating that RepoSummary’s generated features are closer in granularity to manual documentation.

\begin{summarybox}
\textbf{RQ1 Summary:}
RepoSummary achieves substantially higher coverage, maintaining a score of 1.00 in covered and improving by 0.099 in Completely Covered when extracting features from source code, compared with the state-of-the-art baseline (HGEN). Our approach effectively addresses the problem that the features generated by existing approaches are too rough.
\end{summarybox}

\subsection{RQ2. Traceability Link Evaluation } 

We evaluate the effectiveness of traceability links using three metrics: precision, recall, and F1-score ~\cite{Hayes_Dekhtyar_Sundaram_2006}.
Because Claude 3 performed poorly in RQ1 and traceability links depend on RQ1 outcomes, we focus our analysis on the three approaches using GPT-4o mini.

Answering this question helps investigate the accuracy of traceability link (TL) generation.
The results are presented in Table~\ref{tab:traceability link}. We highlight the following observations:
\textbf{Significant recall improvement}. Our approach increases recall by 0.231 over HGEN. HMCS links all files within the related module, which tends to yield higher recall. Nevertheless, on \texttt{Dronology}, the repository with the most complex traceability links, our approach surpasses HMCS by 0.118 in recall.
\textbf{More balanced performance}. In terms of F1, our method improves by 0.025 over HGEN and by 0.189 over HMCS. Our approach balances precision (P) and recall (R), resulting in a higher F1-score compared with the state-of-the-art.

\begin{figure}
    \centering
    \includegraphics[width=0.95\linewidth]{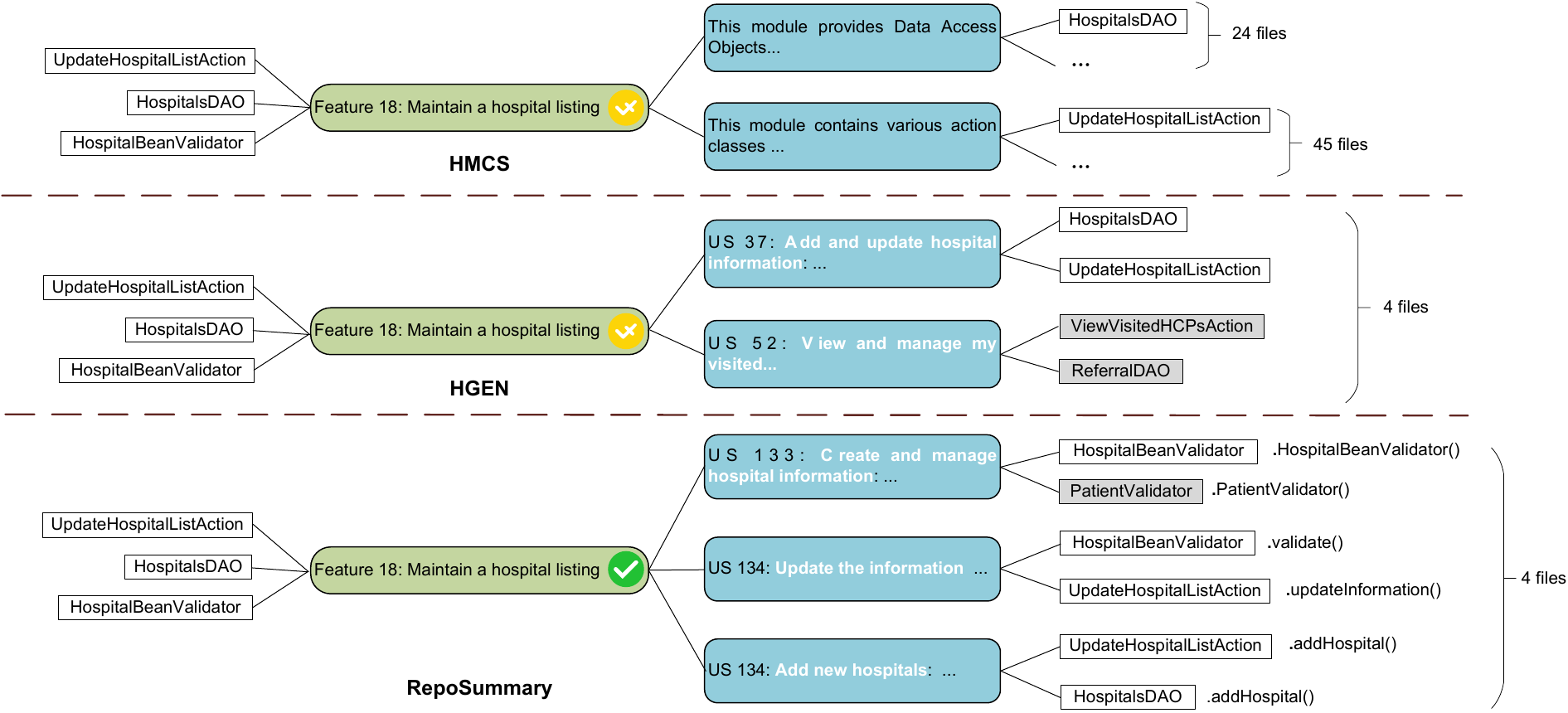}
    \caption{An example of different approaches generated traceability links.}
    \label{fig:RQ2-example}
\end{figure}

During coverage evaluation, we determine the relevance of each ground truth feature (Section \ref{Automatic Evaluation}). Because generated features are derived from file or method level summaries, they naturally induce traceability links. Fig.~\ref{fig:RQ2-example} presents the traceability links for Feature 18 in the \texttt{iTrust} repository across the three approaches. HMCS links each feature to all files in the related module, yielding 69 links with 2 correct. HGEN links to the cluster of files associated with the related features, yielding 4 links with 2 correct. RepoSummary links to the set of files containing methods in the related method clusters, yielding 4 links with 3 correct.

\begin{table}
  \caption{ Traceability link results of different approaches.}
  \label{tab:traceability link}
  \centering
  \small
  \begin{tabular}{L *{9}{C}}
    \toprule
    \multirow{2}{*}{\makecell[l]{\textbf{Approach}}} &
    \multicolumn{3}{c}{Dronology} &
    \multicolumn{3}{c}{eTour} &
    \multicolumn{3}{c}{iTrust} \\
    \cmidrule(lr){2-4} \cmidrule(lr){5-7} \cmidrule(lr){8-10} 
    & P & R & F$_1$
    & P & R & F$_1$
    & P & R & F$_1$\\
    \midrule
    HMCS (GPT-4o mini)
      & .076 & .313 & .122
      & .089 & .536 & .153
      & .076 & \textbf{.780} & .139 \\
    HGEN (GPT-4o mini)
      & \textbf{.148} & .181 & .163
      & \textbf{.404} & .360 & .381
      & .365 & .357 & .361 \\
    RepoSummary (GPT-4o mini)
      & .138 & \textbf{.431} & \textbf{.209}
      & .227 & \textbf{.682} & .340
      & .395 & .478 & \textbf{.433} \\
\hline
    HMCS (Claude-3)
      & .068 & .238  & .106
      & .110 & .539 & .183
      & .077 & .637 & .138 \\
    HGEN (Claude-3)
      & .133 & .165 & .147
      & .356 & .328 & .341
      & \textbf{.428} & .330 & .373 \\
    RepoSummary (Claude-3)
      & .106 & .260 & .151
      & .273 & .656 & \textbf{.386} 
      & .251 & .478 & .330 \\
    \bottomrule
  \end{tabular}
\end{table}
\begin{summarybox}\textbf{RQ2 Summary:}
Compared with existing approaches, our approach significantly improves the recall of traceability links (improves 0.231 over the SOTA) and achieves a higher F1-score. These results indicate that our method aligns more closely with manually extracted feature traceability links.
\end{summarybox}

\subsection{RQ3. Helpfulness Evaluation}
\begin{table}
\caption{Helpfulness evaluation results.}
\label{tab:helpfulness evaluation}
\centering
\small
\begin{tabular}{lccccccccc}
\toprule
\multirow{2}{*}{\textbf{Approach}} & \multicolumn{3}{c}{Link@1} & \multicolumn{3}{c}{Link@2} & \multicolumn{3}{c}{Link@3} \\
\cmidrule(lr){2-4}\cmidrule(lr){5-7}\cmidrule(lr){8-10}
 & P & R & F$_1$ & P & R & F$_1$ & P & R & F$_1$ \\
\midrule
HMCS & .049 & .234 & .081 & .034 & .402 & .062 & .032 & .493 & .060\\
HGEN & .194 & .171 & .182 & .130 & .238 & .168 & .128 & .353 & .188\\
RepoSummary & \textbf{.331} & \textbf{.350} & \textbf{.340} & \textbf{.314} & \textbf{.563} & \textbf{.403} & \textbf{.301} & \textbf{.622} & \textbf{.406}\\

\bottomrule
\end{tabular}
\end{table}

We operationalize developer helpfulness as observable evaluation objectives: navigation and localization capability \cite{yang2025docagent}. Concretely, we assess whether the documentation guides developers more quickly, more accurately, and more comprehensively to methods relevant to the target functionality.

To assess the helpfulness of software documentation in realistic development settings, we curated 26 commits and their corresponding commit messages from three GitHub projects.
Each commit message is a short, author-provided summary describing the change; the data format is illustrated in Fig.~\ref{fig:commit-complexity}. We embedded the commit messages using Sentence-BERT and computed their similarities to the features produced by the three approaches. For each approach, we selected the top three most similar features to compute Link@k, and evaluated precision (P), recall (R), and F1-score based on whether the links associated with the top-k generated features hit the methods modified in the commit.

Because HMCS and HGEN link only at the file level, we treat all methods within a linked file as related to the generated feature for evaluation.

RepoSummary enables developers to locate the relevant methods more accurately and comprehensively, given a development task. Notably, HMCS and HGEN include all methods in a file for evaluation and, in principle, should achieve high recall. However, RepoSummary consistently attains the highest recall, peaking at 0.622. This suggests that refining the analysis granularity to the method level can better identify file functionality and organize it into the correct feature.

As shown in Fig.~\ref{fig:commit-example}, the development task “Backend Management—Complete CRUD operations for tags” spans six Java files and involves nine methods connected by various relationships (e.g., calls and implementations). Without consulting the results of RepoSummary, a developer must manually locate each relevant file and method and understand the underlying code before proceeding. In contrast, with our approach, developers can quickly identify the methods related to the task and use the feature flow to grasp the execution logic, thereby reducing cognitive load.

\begin{summarybox}
    \textbf{RQ3 Summary:} In real-world development settings, RepoSummary can help developers more accurately and comprehensively locate relevant methods and reduce the cognitive load associated with manual code navigation and comprehension.
\end{summarybox}

\subsection{RQ4. Ablation Study } 
\begin{table}
  \centering  
  \caption{The ablation results.}
  \label{tab:Ablation study}
  \renewcommand{\arraystretch}{0.8}
  \setlength{\tabcolsep}{5pt}
  \small
  \begin{tabular}{l l r r r r r r }
    \toprule
    \multicolumn{2}{l}{\textbf{Variant}} & \textbf{C} & \textbf{CB} & \textbf{CC}& \textbf{P} & \textbf{R}& \textbf{F$_1$} \\
    \midrule
    \multirow{3}{*}{RepoSummary}
      & avg   & 1.00 & .587 & .711 & .253 & .530 & .327 \\
      & v.s.\ RepoSummary & /  & /  & /  & / & /& /\\
      & v.s.\ HGEN  & ↑3.0\% & ↓4.0\% & ↑16.2\% & ↓17.1\% &↑77.2\% & ↑8.5\% \\
    \midrule
    \midrule
    \multirow{3}{*}{w/o Adj}
      & avg               & .997 & .702 & .611 & .173 & .475& .252\\
      & v.s.\ RepoSummary & ↓3.0\% & ↑19.7\% &  ↓14.1\% & ↓31.8\% &↓10.4\% & ↓22.9\% \\
      & v.s.\ HGEN    & 0.0\% & ↑14.9\% & ↓1.0\% & ↓43.5\% & ↑58.8\% &  ↓16.4\%\\
    \midrule
    \multirow{3}{*}{w/o SS}
      & avg               & .980 & .671 & .517 & .180 &.389 & .237\\
      & v.s.\ RepoSummary & ↓2.0\% & ↑14.3\% & ↓27.2\% & ↓29.1\% & ↓26.7\% & ↓27.5\% \\
      & v.s.\ HGEN & ↓1.7\% & ↑9.8\% & ↓15.4\% & ↓41.2\% & ↑29.8\% & ↓21.3\% \\
    \midrule
    \midrule
    \multirow{3}{*}{w/o FLC}
      & avg               & .984 & .690 & .588 &.146 & .544 & .230 \\
      & v.s.\ RepoSummary & ↓1.6\% & ↑17.6\% & ↓17.3\% & ↓42.2\% & ↑2.5\% & ↓29.8\%\\
      & v.s.\ HGEN  & ↓1.3\% & ↑12.9\% & ↓3.9\%  & ↓52.1\% & ↑81.6\% & ↓23.9\%\\
    \midrule
    \multirow{3}{*}{w/o MLC}
      & avg               & .969 & .789 & .385 & .135& .369&.184 \\
      & v.s.\ RepoSummary & ↓3.1\% & ↑34.4\% & ↓45.9\% & ↓46.8\% & ↓30.4\% & ↓43.7\% \\
      & v.s.\ HGEN        & ↓2.8\% & ↑29.1\% & ↓3.71\% & ↓55.9\% & ↑23.4\% & ↓38.9\%\\

    \bottomrule
  \end{tabular}
\end{table}
 To evaluate the contribution of each component in our
approach, we conduct an ablation study examining: (1) the impact of different types of matrices in clustering; (2) the impact of different levels of clustering. So we set up four variant approaches. 
\begin{itemize}
\item \textbf{w/o Adj} indicates that adjacency matrices are not used during hierarchical clustering.

\item \textbf{w/o SS} indicates that semantic similarity matrices are not used during hierarchical clustering.

\item \textbf{w/o MLC} indicates clustering at the file level only. This variant focuses on file-level code clustering by analyzing the dependencies between files and the semantic similarity information of the file summaries. And this variant is often used in architecture refactoring or modular analysis scenarios ~\cite{zhang2023software}. 

\item \textbf{w/o FLC} indicates clustering at the method level only. This variant focuses on the method-level clustering and establishes fine-grained associations by analyzing the dependencies between methods and the semantic similarity information of method summaries, ignoring the physical location of the file level. This method-level focus is especially useful for scenarios of identifying code smells ~\cite{10.1145/3387906.3388625}.
\end{itemize}

Compared with the complete approach, the four variant approaches have a decrease in feature coverage and traceability link at the overall level, but the lack of method-level clustering has the greatest impact on the complete approach. Only when clustering at the method level, the recall of the traceability link is increased by 2.5\% compared to the complete approach. It shows that the method-level cluster is very effective to improve the recall of traceability link.

\begin{summarybox}
\textbf{RQ4 Summary:} Each component contributes meaningfully to overall performance. SS weights are more important in multi-weight clustering. And method-level clustering is very important in hierarchical clustering.
\end{summarybox}

\section{Discussion}
\label{sec:discussion}

In this section, we discuss the limitations and future work, as well as the threats to validity.

\subsection{Limitations and Future Work}

This section discusses the limitations of our current work and the potential directions for future research.

\textbf{Limitations in Programming Language Types}: The structural analysis of the code repository introduces constraints tied to specific programming languages. While it is theoretically feasible to replace the static analysis tool with one capable of summarizing repositories in other languages, our current implementation integrates support only for Java. We plan to extend our approach to other programming languages in future work.

\textbf{Limitations in Time Efficiency for Large Code Repository}: In our approach, Step 2 (hierarchical code summarization) and 4 (repository summarization) require the integration of LLMs, resulting in significant time overhead. However, these steps can be executed in parallel since the order of method calls is not considered during summarization or feature generation. Future improvements will focus on enhancing the time efficiency of our approach through optimized parallel processing.

\subsection{Threats to Validity}
\label{sec:threats to validity}

Our experiment results may suffer from several threats to
validity. We discuss internal validity and external validity respectively.

\subsubsection{Internal Validity}
The threats to internal validity include:
\textbf{1) Dataset Selection}. We conduct the feature coverage and traceability link evaluation on three datasets. The limited number of datasets could potentially affect the generalizability of evaluation results. However, these datasets cover different domains, which helps to mitigate the risk of domain-specific biases. Besides these three datasets, we also collect commit datasets from GitHub repositories to better simulate real-world development scenarios, ensuring that our approach is validated in the context of actual software development. 
\textbf{2) Prompt Design}. We use LLMs in different steps in our approach, including hierarchical code summarization, repository summarization, and feature coverage evaluation. The effectiveness of our approach may be influenced by the choice of prompts. Due to budget limitations, the prompts explored in this paper are currently limited. They might not fully unleash models’ capabilities. Therefore, we will explore more effective prompt strategies in the future to improve our generated results.
\textbf{3) Automatic Evaluation Framework}. For feature coverage evaluation, we discuss the concerns of conventional evaluation metrics and design a two-stage automatic evaluation framework based on LLM-as-a-judge. The reliability of using LLMs as evaluators remains a potential threat. To mitigate this threat, we compare the evaluation results with human evaluations and find a strong correlation between them, suggesting the reliability of our framework.

\subsubsection{External Validity}
We use two kinds of datasets to conduct evaluation, repository datasets and commit datasets, thereby addressing concerns about the generalizability of results. However, these datasets are limited to open-source projects written in the Java language, which constrains the applicability of our conclusions to other programming languages or closed-source projects. In fact, our approach can be adapted to other languages. By replacing the static analysis tool we used with equivalent tools designed for other programming languages, our approach can be extended to extract file nodes, method nodes, and their mutual relationships from software projects written in other languages, such as Python, C++.

Another potential threat to generalizability lies in the use of a specific LLM. To address this, we conduct experiments on two different LLMs, GPT-4o mini and Claude 3. The selection of LLMs might influence the experimental results. For instance, using smaller or less capable LLMs could lead to different experimental conclusions. Therefore, further exploration with a wider variety of LLMs, including open-source alternatives, is needed to better assess the robustness of our approach.

\section{Conclusion}
\label{sec:conclusion}
In this paper, we have proposed RepoSummary, a feature-oriented code repository summarization approach that simultaneously generates repository documentation.
By integrating hierarchical summarization and hierarchical clustering, our approach enables a more precise feature-to-code traceability link than existing approaches.
Evaluated on three repositories and 26 commits from a real development environment, our approach demonstrates higher coverage and more accurate traceability, and offers greater practical utility for repository comprehension. 
Consequently, our proposed approach can significantly assist in software maintenance and reuse tasks, while also providing valuable guidance for Traceability Link Recovery (TLR) work.
In future work, we plan to extend our approach to additional programming languages and evaluate it on larger and more diverse datasets.

\bibliographystyle{ACM-Reference-Format}
\bibliography{reference}


\begin{thebibliography}{54}


\ifx \showCODEN    \undefined \def \showCODEN     #1{\unskip}     \fi
\ifx \showISBNx    \undefined \def \showISBNx     #1{\unskip}     \fi
\ifx \showISBNxiii \undefined \def \showISBNxiii  #1{\unskip}     \fi
\ifx \showISSN     \undefined \def \showISSN      #1{\unskip}     \fi
\ifx \showLCCN     \undefined \def \showLCCN      #1{\unskip}     \fi
\ifx \shownote     \undefined \def \shownote      #1{#1}          \fi
\ifx \showarticletitle \undefined \def \showarticletitle #1{#1}   \fi
\ifx \showURL      \undefined \def \showURL       {\relax}        \fi
\providecommand\bibfield[2]{#2}
\providecommand\bibinfo[2]{#2}
\providecommand\natexlab[1]{#1}
\providecommand\showeprint[2][]{arXiv:#2}

\bibitem[gpt({[n.\,d.]})]%
        {gpt-4o}
 \bibinfo{year}{[n.\,d.]}\natexlab{}.
\newblock
  \bibinfo{howpublished}{\url{https://platform.openai.com/docs/models/gpt-4o}}.
\newblock


\bibitem[Ahmad et~al\mbox{.}(2020)]%
        {ahmad2020transformer}
\bibfield{author}{\bibinfo{person}{Wasi~Uddin Ahmad}, \bibinfo{person}{Saikat
  Chakraborty}, \bibinfo{person}{Baishakhi Ray}, {and} \bibinfo{person}{Kai-Wei
  Chang}.} \bibinfo{year}{2020}\natexlab{}.
\newblock \showarticletitle{A transformer-based approach for source code
  summarization}.
\newblock \bibinfo{journal}{\emph{arXiv preprint arXiv:2005.00653}}
  (\bibinfo{year}{2020}).
\newblock


\bibitem[Bansal et~al\mbox{.}(2023)]%
        {bansal2023function}
\bibfield{author}{\bibinfo{person}{Aakash Bansal}, \bibinfo{person}{Zachary
  Eberhart}, \bibinfo{person}{Zachary Karas}, \bibinfo{person}{Yu Huang}, {and}
  \bibinfo{person}{Collin McMillan}.} \bibinfo{year}{2023}\natexlab{}.
\newblock \showarticletitle{Function call graph context encoding for neural
  source code summarization}.
\newblock \bibinfo{journal}{\emph{IEEE Transactions on Software Engineering}}
  \bibinfo{volume}{49}, \bibinfo{number}{9} (\bibinfo{year}{2023}),
  \bibinfo{pages}{4268--4281}.
\newblock


\bibitem[Chang et~al\mbox{.}(2023)]%
        {chang2023booookscore}
\bibfield{author}{\bibinfo{person}{Yapei Chang}, \bibinfo{person}{Kyle Lo},
  \bibinfo{person}{Tanya Goyal}, {and} \bibinfo{person}{Mohit Iyyer}.}
  \bibinfo{year}{2023}\natexlab{}.
\newblock \showarticletitle{BooookScore: A systematic exploration of
  book-length summarization in the era of LLMs}. In
  \bibinfo{booktitle}{\emph{International Conference on Learning
  Representations}}.
\newblock


\bibitem[Cleland-Huang et~al\mbox{.}(2018)]%
        {cleland2018dronology}
\bibfield{author}{\bibinfo{person}{Jane Cleland-Huang},
  \bibinfo{person}{Michael Vierhauser}, {and} \bibinfo{person}{Sean Bayley}.}
  \bibinfo{year}{2018}\natexlab{}.
\newblock \showarticletitle{Dronology: An incubator for cyber-physical system
  research}.
\newblock \bibinfo{journal}{\emph{arXiv preprint arXiv:1804.02423}}
  (\bibinfo{year}{2018}).
\newblock


\bibitem[de~Souza et~al\mbox{.}(2005)]%
        {sigdoc/SouzaAO05}
\bibfield{author}{\bibinfo{person}{Sergio Cozzetti~B. de Souza},
  \bibinfo{person}{Nicolas Anquetil}, {and}
  \bibinfo{person}{K{\'{a}}thia~Mar{\c{c}}al de Oliveira}.}
  \bibinfo{year}{2005}\natexlab{}.
\newblock \showarticletitle{A study of the documentation essential to software
  maintenance}. In \bibinfo{booktitle}{\emph{Proceedings of the 23rd Annual
  International Conference on Design of Communication: documenting {\&}
  Designing for Pervasive Information, {SIGDOC} 2005, Coventry, UK, September
  21-23, 2005}}, \bibfield{editor}{\bibinfo{person}{Scott~R. Tilley} {and}
  \bibinfo{person}{Robert~M. Newman}} (Eds.). \bibinfo{publisher}{{ACM}},
  \bibinfo{pages}{68--75}.
\newblock
\href{https://doi.org/10.1145/1085313.1085331}{doi:\nolinkurl{10.1145/1085313.1085331}}


\bibitem[Dearstyne et~al\mbox{.}(2024)]%
        {dearstyne2024supporting}
\bibfield{author}{\bibinfo{person}{Katherine~R Dearstyne},
  \bibinfo{person}{Alberto~D Rodriguez}, {and} \bibinfo{person}{Jane
  Cleland-Huang}.} \bibinfo{year}{2024}\natexlab{}.
\newblock \showarticletitle{Supporting Software Maintenance with Dynamically
  Generated Document Hierarchies}. In \bibinfo{booktitle}{\emph{2024 IEEE
  International Conference on Software Maintenance and Evolution (ICSME)}}.
  IEEE, \bibinfo{pages}{426--437}.
\newblock


\bibitem[Dhulshette et~al\mbox{.}(2025)]%
        {dhulshette2025hierarchical}
\bibfield{author}{\bibinfo{person}{Nilesh Dhulshette}, \bibinfo{person}{Sapan
  Shah}, {and} \bibinfo{person}{Vinay Kulkarni}.}
  \bibinfo{year}{2025}\natexlab{}.
\newblock \showarticletitle{Hierarchical Repository-Level Code Summarization
  for Business Applications Using Local LLMs}.
\newblock \bibinfo{journal}{\emph{arXiv preprint arXiv:2501.07857}}
  (\bibinfo{year}{2025}).
\newblock


\bibitem[Feng et~al\mbox{.}(2020)]%
        {feng-etal-2020-codebert}
\bibfield{author}{\bibinfo{person}{Zhangyin Feng}, \bibinfo{person}{Daya Guo},
  \bibinfo{person}{Duyu Tang}, \bibinfo{person}{Nan Duan},
  \bibinfo{person}{Xiaocheng Feng}, \bibinfo{person}{Ming Gong},
  \bibinfo{person}{Linjun Shou}, \bibinfo{person}{Bing Qin},
  \bibinfo{person}{Ting Liu}, \bibinfo{person}{Daxin Jiang}, {and}
  \bibinfo{person}{Ming Zhou}.} \bibinfo{year}{2020}\natexlab{}.
\newblock \showarticletitle{{C}ode{BERT}: A Pre-Trained Model for Programming
  and Natural Languages}. In \bibinfo{booktitle}{\emph{Findings of the
  Association for Computational Linguistics: EMNLP 2020}}.
  \bibinfo{publisher}{Association for Computational Linguistics},
  \bibinfo{address}{Online}, \bibinfo{pages}{1536--1547}.
\newblock
\href{https://doi.org/10.18653/v1/2020.findings-emnlp.139}{doi:\nolinkurl{10.18653/v1/2020.findings-emnlp.139}}


\bibitem[Fuch{\ss} et~al\mbox{.}(2025)]%
        {fuchss2025lissa}
\bibfield{author}{\bibinfo{person}{Dominik Fuch{\ss}}, \bibinfo{person}{Tobias
  Hey}, \bibinfo{person}{Jan Keim}, \bibinfo{person}{Haoyu Liu},
  \bibinfo{person}{Niklas Ewald}, \bibinfo{person}{Tobias Thirolf}, {and}
  \bibinfo{person}{Anne Koziolek}.} \bibinfo{year}{2025}\natexlab{}.
\newblock \showarticletitle{LiSSA: toward generic traceability link recovery
  through retrieval-augmented generation}. In
  \bibinfo{booktitle}{\emph{Proceedings of the IEEE/ACM 47th International
  Conference on Software Engineering. ICSE}}, Vol.~\bibinfo{volume}{25}.
\newblock


\bibitem[Gamma et~al\mbox{.}(1995)]%
        {gamma1995design}
\bibfield{author}{\bibinfo{person}{Erich Gamma}, \bibinfo{person}{Richard
  Helm}, \bibinfo{person}{Ralph Johnson}, {and} \bibinfo{person}{John
  Vlissides}.} \bibinfo{year}{1995}\natexlab{}.
\newblock \bibinfo{booktitle}{\emph{Design patterns: elements of reusable
  object-oriented software}}.
\newblock \bibinfo{publisher}{Pearson Deutschland GmbH}.
\newblock


\bibitem[Gao et~al\mbox{.}(2021)]%
        {Gao_Yao_Chen_2021}
\bibfield{author}{\bibinfo{person}{Tianyu Gao}, \bibinfo{person}{Xingcheng
  Yao}, {and} \bibinfo{person}{Danqi Chen}.} \bibinfo{year}{2021}\natexlab{}.
\newblock \showarticletitle{SimCSE: Simple Contrastive Learning of Sentence
  Embeddings}. In \bibinfo{booktitle}{\emph{Proceedings of the 2021 Conference
  on Empirical Methods in Natural Language Processing}}.
\newblock
\href{https://doi.org/10.18653/v1/2021.emnlp-main.552}{doi:\nolinkurl{10.18653/v1/2021.emnlp-main.552}}


\bibitem[Geng et~al\mbox{.}(2024)]%
        {10.1145/3597503.3608134}
\bibfield{author}{\bibinfo{person}{Mingyang Geng}, \bibinfo{person}{Shangwen
  Wang}, \bibinfo{person}{Dezun Dong}, \bibinfo{person}{Haotian Wang},
  \bibinfo{person}{Ge Li}, \bibinfo{person}{Zhi Jin},
  \bibinfo{person}{Xiaoguang Mao}, {and} \bibinfo{person}{Xiangke Liao}.}
  \bibinfo{year}{2024}\natexlab{}.
\newblock \showarticletitle{Large Language Models are Few-Shot Summarizers:
  Multi-Intent Comment Generation via In-Context Learning}. In
  \bibinfo{booktitle}{\emph{Proceedings of the 46th IEEE/ACM International
  Conference on Software Engineering}} (, Lisbon, Portugal,)
  \emph{(\bibinfo{series}{ICSE '24})}. \bibinfo{publisher}{Association for
  Computing Machinery}, \bibinfo{address}{New York, NY, USA}, Article
  \bibinfo{articleno}{39}, \bibinfo{numpages}{13}~pages.
\newblock
\showISBNx{9798400702174}
\href{https://doi.org/10.1145/3597503.3608134}{doi:\nolinkurl{10.1145/3597503.3608134}}


\bibitem[Hayes et~al\mbox{.}(2006)]%
        {Hayes_Dekhtyar_Sundaram_2006}
\bibfield{author}{\bibinfo{person}{J.H. Hayes}, \bibinfo{person}{A. Dekhtyar},
  {and} \bibinfo{person}{S.K. Sundaram}.} \bibinfo{year}{2006}\natexlab{}.
\newblock \showarticletitle{Advancing candidate link generation for
  requirements tracing: the study of methods}.
\newblock \bibinfo{journal}{\emph{IEEE Transactions on Software Engineering}}
  \bibinfo{volume}{32}, \bibinfo{number}{1} (\bibinfo{date}{Jan}
  \bibinfo{year}{2006}), \bibinfo{pages}{4–19}.
\newblock
\href{https://doi.org/10.1109/tse.2006.3}{doi:\nolinkurl{10.1109/tse.2006.3}}


\bibitem[Hu et~al\mbox{.}(2018)]%
        {hu2018deep}
\bibfield{author}{\bibinfo{person}{Xing Hu}, \bibinfo{person}{Ge Li},
  \bibinfo{person}{Xin Xia}, \bibinfo{person}{David Lo}, {and}
  \bibinfo{person}{Zhi Jin}.} \bibinfo{year}{2018}\natexlab{}.
\newblock \showarticletitle{Deep code comment generation}. In
  \bibinfo{booktitle}{\emph{2018 IEEE/ACM 26th International Conference on
  Program Comprehension (ICPC)}}. IEEE, \bibinfo{pages}{200--20010}.
\newblock


\bibitem[Hu et~al\mbox{.}(2022)]%
        {hu2022practitioners}
\bibfield{author}{\bibinfo{person}{Xing Hu}, \bibinfo{person}{Xin Xia},
  \bibinfo{person}{David Lo}, \bibinfo{person}{Zhiyuan Wan},
  \bibinfo{person}{Qiuyuan Chen}, {and} \bibinfo{person}{Thomas Zimmermann}.}
  \bibinfo{year}{2022}\natexlab{}.
\newblock \showarticletitle{Practitioners' expectations on automated code
  comment generation}. In \bibinfo{booktitle}{\emph{Proceedings of the 44th
  international conference on software engineering}}.
  \bibinfo{pages}{1693--1705}.
\newblock


\bibitem[Iyer et~al\mbox{.}(2016)]%
        {iyer2016summarizing}
\bibfield{author}{\bibinfo{person}{Srinivasan Iyer}, \bibinfo{person}{Ioannis
  Konstas}, \bibinfo{person}{Alvin Cheung}, {and} \bibinfo{person}{Luke
  Zettlemoyer}.} \bibinfo{year}{2016}\natexlab{}.
\newblock \showarticletitle{Summarizing source code using a neural attention
  model}. In \bibinfo{booktitle}{\emph{Proceedings of the 54th Annual Meeting
  of the Association for Computational Linguistics (Volume 1: Long Papers)}}.
  \bibinfo{pages}{2073--2083}.
\newblock


\bibitem[Khan and Uddin(2023)]%
        {10.1145/3551349.3559548}
\bibfield{author}{\bibinfo{person}{Junaed~Younus Khan} {and}
  \bibinfo{person}{Gias Uddin}.} \bibinfo{year}{2023}\natexlab{}.
\newblock \showarticletitle{Automatic Code Documentation Generation Using
  GPT-3}. In \bibinfo{booktitle}{\emph{Proceedings of the 37th IEEE/ACM
  International Conference on Automated Software Engineering}} (, Rochester,
  MI, USA,) \emph{(\bibinfo{series}{ASE '22})}. \bibinfo{publisher}{Association
  for Computing Machinery}, \bibinfo{address}{New York, NY, USA}, Article
  \bibinfo{articleno}{174}, \bibinfo{numpages}{6}~pages.
\newblock
\showISBNx{9781450394758}
\href{https://doi.org/10.1145/3551349.3559548}{doi:\nolinkurl{10.1145/3551349.3559548}}


\bibitem[LeClair et~al\mbox{.}(2019)]%
        {leclair2019neural}
\bibfield{author}{\bibinfo{person}{Alexander LeClair}, \bibinfo{person}{Siyuan
  Jiang}, {and} \bibinfo{person}{Collin McMillan}.}
  \bibinfo{year}{2019}\natexlab{}.
\newblock \showarticletitle{A neural model for generating natural language
  summaries of program subroutines}. In \bibinfo{booktitle}{\emph{2019 IEEE/ACM
  41st International Conference on Software Engineering (ICSE)}}. IEEE,
  \bibinfo{pages}{795--806}.
\newblock


\bibitem[Li et~al\mbox{.}(2025)]%
        {li2025generation}
\bibfield{author}{\bibinfo{person}{Dawei Li}, \bibinfo{person}{Bohan Jiang},
  \bibinfo{person}{Liangjie Huang}, \bibinfo{person}{Alimohammad Beigi},
  \bibinfo{person}{Chengshuai Zhao}, \bibinfo{person}{Zhen Tan},
  \bibinfo{person}{Amrita Bhattacharjee}, \bibinfo{person}{Yuxuan Jiang},
  \bibinfo{person}{Canyu Chen}, \bibinfo{person}{Tianhao Wu}, {et~al\mbox{.}}}
  \bibinfo{year}{2025}\natexlab{}.
\newblock \showarticletitle{From generation to judgment: Opportunities and
  challenges of llm-as-a-judge, 2025}.
\newblock \bibinfo{journal}{\emph{URL https://arxiv. org/abs/2411.16594}}
  (\bibinfo{year}{2025}).
\newblock


\bibitem[Li et~al\mbox{.}(2021)]%
        {9678724}
\bibfield{author}{\bibinfo{person}{Jia Li}, \bibinfo{person}{Yongmin Li},
  \bibinfo{person}{Ge Li}, \bibinfo{person}{Xing Hu}, \bibinfo{person}{Xin
  Xia}, {and} \bibinfo{person}{Zhi Jin}.} \bibinfo{year}{2021}\natexlab{}.
\newblock \showarticletitle{EditSum: A Retrieve-and-Edit Framework for Source
  Code Summarization}. In \bibinfo{booktitle}{\emph{2021 36th IEEE/ACM
  International Conference on Automated Software Engineering (ASE)}}.
  \bibinfo{pages}{155--166}.
\newblock
\href{https://doi.org/10.1109/ASE51524.2021.9678724}{doi:\nolinkurl{10.1109/ASE51524.2021.9678724}}


\bibitem[Lin and Chen({[n.\,d.]})]%
        {Lin_Chen}
\bibfield{author}{\bibinfo{person}{Yen-Ting Lin} {and}
  \bibinfo{person}{Yun-Nung Chen}.} \bibinfo{year}{[n.\,d.]}\natexlab{}.
\newblock \showarticletitle{LLM-EVAL: Unified Multi-Dimensional Automatic
  Evaluation for Open-Domain Conversations with Large Language Models}.
\newblock  (\bibinfo{year}{[n.\,d.]}).
\newblock


\bibitem[Liu et~al\mbox{.}(2019)]%
        {10.1145/3361242.3362774}
\bibfield{author}{\bibinfo{person}{Bohong Liu}, \bibinfo{person}{Tao Wang},
  \bibinfo{person}{Xunhui Zhang}, \bibinfo{person}{Qiang Fan},
  \bibinfo{person}{Gang Yin}, {and} \bibinfo{person}{Jinsheng Deng}.}
  \bibinfo{year}{2019}\natexlab{}.
\newblock \showarticletitle{A Neural-Network Based Code Summarization Approach
  by Using Source Code and Its Call Dependencies}. In
  \bibinfo{booktitle}{\emph{Proceedings of the 11th Asia-Pacific Symposium on
  Internetware}} (Fukuoka, Japan) \emph{(\bibinfo{series}{Internetware '19})}.
  \bibinfo{publisher}{Association for Computing Machinery},
  \bibinfo{address}{New York, NY, USA}, Article \bibinfo{articleno}{12},
  \bibinfo{numpages}{10}~pages.
\newblock
\showISBNx{9781450377010}
\href{https://doi.org/10.1145/3361242.3362774}{doi:\nolinkurl{10.1145/3361242.3362774}}


\bibitem[Lomshakov et~al\mbox{.}(2024)]%
        {emnlp/LomshakovPSBLN24}
\bibfield{author}{\bibinfo{person}{Vadim Lomshakov}, \bibinfo{person}{Andrey
  Podivilov}, \bibinfo{person}{Sergey Savin}, \bibinfo{person}{Oleg
  Baryshnikov}, \bibinfo{person}{Alena Lisevych}, {and}
  \bibinfo{person}{Sergey~I. Nikolenko}.} \bibinfo{year}{2024}\natexlab{}.
\newblock \showarticletitle{ProConSuL: Project Context for Code Summarization
  with LLMs}. In \bibinfo{booktitle}{\emph{Proceedings of the 2024 Conference
  on Empirical Methods in Natural Language Processing: {EMNLP} 2024 - Industry
  Track, Miami, Florida, USA, November 12-16, 2024}},
  \bibfield{editor}{\bibinfo{person}{Franck Dernoncourt},
  \bibinfo{person}{Daniel Preotiuc{-}Pietro}, {and} \bibinfo{person}{Anastasia
  Shimorina}} (Eds.). \bibinfo{publisher}{Association for Computational
  Linguistics}, \bibinfo{pages}{866--880}.
\newblock
\urldef\tempurl%
\url{https://aclanthology.org/2024.emnlp-industry.65}
\showURL{%
\tempurl}


\bibitem[Lu and Liu(2024)]%
        {lu2024improving}
\bibfield{author}{\bibinfo{person}{Hanzhen Lu} {and} \bibinfo{person}{Zhongxin
  Liu}.} \bibinfo{year}{2024}\natexlab{}.
\newblock \showarticletitle{Improving Retrieval-Augmented Code Comment
  Generation by Retrieving for Generation}. In \bibinfo{booktitle}{\emph{2024
  IEEE International Conference on Software Maintenance and Evolution
  (ICSME)}}. IEEE, \bibinfo{pages}{350--362}.
\newblock


\bibitem[Luo et~al\mbox{.}(2024)]%
        {luo2024repoagent}
\bibfield{author}{\bibinfo{person}{Qinyu Luo}, \bibinfo{person}{Yining Ye},
  \bibinfo{person}{Shihao Liang}, \bibinfo{person}{Zhong Zhang},
  \bibinfo{person}{Yujia Qin}, \bibinfo{person}{Yaxi Lu},
  \bibinfo{person}{Yesai Wu}, \bibinfo{person}{Xin Cong},
  \bibinfo{person}{Yankai Lin}, \bibinfo{person}{Yingli Zhang},
  {et~al\mbox{.}}} \bibinfo{year}{2024}\natexlab{}.
\newblock \showarticletitle{RepoAgent: An LLM-Powered Open-Source Framework for
  Repository-level Code Documentation Generation}. In
  \bibinfo{booktitle}{\emph{Proceedings of the 2024 Conference on Empirical
  Methods in Natural Language Processing: System Demonstrations}}.
  \bibinfo{pages}{436--464}.
\newblock


\bibitem[Ma et~al\mbox{.}(2024)]%
        {ma2024alibaba}
\bibfield{author}{\bibinfo{person}{Yingwei Ma}, \bibinfo{person}{Qingping
  Yang}, \bibinfo{person}{Rongyu Cao}, \bibinfo{person}{Binhua Li},
  \bibinfo{person}{Fei Huang}, {and} \bibinfo{person}{Yongbin Li}.}
  \bibinfo{year}{2024}\natexlab{}.
\newblock \showarticletitle{Alibaba LingmaAgent: Improving Automated Issue
  Resolution via Comprehensive Repository Exploration}.
\newblock \bibinfo{journal}{\emph{arXiv preprint arXiv:2406.01422}}
  (\bibinfo{year}{2024}).
\newblock


\bibitem[McBurney et~al\mbox{.}(2018)]%
        {tse/McBurneyJKKAMM18}
\bibfield{author}{\bibinfo{person}{Paul~W. McBurney}, \bibinfo{person}{Siyuan
  Jiang}, \bibinfo{person}{Marouane Kessentini}, \bibinfo{person}{Nicholas~A.
  Kraft}, \bibinfo{person}{Ameer Armaly}, \bibinfo{person}{Mohamed~Wiem
  Mkaouer}, {and} \bibinfo{person}{Collin McMillan}.}
  \bibinfo{year}{2018}\natexlab{}.
\newblock \showarticletitle{Towards Prioritizing Documentation Effort}.
\newblock \bibinfo{journal}{\emph{{IEEE} Trans. Software Eng.}}
  \bibinfo{volume}{44}, \bibinfo{number}{9} (\bibinfo{year}{2018}),
  \bibinfo{pages}{897--913}.
\newblock
\href{https://doi.org/10.1109/TSE.2017.2716950}{doi:\nolinkurl{10.1109/TSE.2017.2716950}}


\bibitem[Naimi et~al\mbox{.}(2024)]%
        {naimi2024automating}
\bibfield{author}{\bibinfo{person}{Lahbib Naimi}, \bibinfo{person}{Abdeslam
  Jakimi}, \bibinfo{person}{Rachid Saadane}, \bibinfo{person}{Abdellah Chehri},
  {et~al\mbox{.}}} \bibinfo{year}{2024}\natexlab{}.
\newblock \showarticletitle{Automating software documentation: Employing llms
  for precise use case description}.
\newblock \bibinfo{journal}{\emph{Procedia Computer Science}}
  \bibinfo{volume}{246} (\bibinfo{year}{2024}), \bibinfo{pages}{1346--1354}.
\newblock


\bibitem[Nybom et~al\mbox{.}(2018)]%
        {nybom2018systematic}
\bibfield{author}{\bibinfo{person}{Kristian Nybom}, \bibinfo{person}{Adnan
  Ashraf}, {and} \bibinfo{person}{Ivan Porres}.}
  \bibinfo{year}{2018}\natexlab{}.
\newblock \showarticletitle{A systematic mapping study on API documentation
  generation approaches}. In \bibinfo{booktitle}{\emph{2018 44th euromicro
  conference on software engineering and advanced applications (SEAA)}}. IEEE,
  \bibinfo{pages}{462--469}.
\newblock


\bibitem[Panichella et~al\mbox{.}(2012)]%
        {panichella2012mining}
\bibfield{author}{\bibinfo{person}{Sebastiano Panichella},
  \bibinfo{person}{Jairo Aponte}, \bibinfo{person}{Massimiliano Di~Penta},
  \bibinfo{person}{Andrian Marcus}, {and} \bibinfo{person}{Gerardo Canfora}.}
  \bibinfo{year}{2012}\natexlab{}.
\newblock \showarticletitle{Mining source code descriptions from developer
  communications}. In \bibinfo{booktitle}{\emph{2012 20th IEEE International
  Conference on Program Comprehension (ICPC)}}. IEEE, \bibinfo{pages}{63--72}.
\newblock


\bibitem[Papineni et~al\mbox{.}(2002)]%
        {bleu}
\bibfield{author}{\bibinfo{person}{Kishore Papineni}, \bibinfo{person}{Salim
  Roukos}, \bibinfo{person}{Todd Ward}, {and} \bibinfo{person}{Wei-Jing Zhu}.}
  \bibinfo{year}{2002}\natexlab{}.
\newblock \showarticletitle{Bleu: a method for automatic evaluation of machine
  translation}. In \bibinfo{booktitle}{\emph{Proceedings of the 40th annual
  meeting of the Association for Computational Linguistics}}.
  \bibinfo{pages}{311--318}.
\newblock


\bibitem[Pigazzini et~al\mbox{.}(2020)]%
        {10.1145/3387906.3388625}
\bibfield{author}{\bibinfo{person}{Ilaria Pigazzini},
  \bibinfo{person}{Francesca~Arcelli Fontana}, \bibinfo{person}{Valentina
  Lenarduzzi}, {and} \bibinfo{person}{Davide Taibi}.}
  \bibinfo{year}{2020}\natexlab{}.
\newblock \showarticletitle{Towards microservice smells detection}. In
  \bibinfo{booktitle}{\emph{Proceedings of the 3rd International Conference on
  Technical Debt}} (Seoul, Republic of Korea) \emph{(\bibinfo{series}{TechDebt
  '20})}. \bibinfo{publisher}{Association for Computing Machinery},
  \bibinfo{address}{New York, NY, USA}, \bibinfo{pages}{92–97}.
\newblock
\showISBNx{9781450379601}
\href{https://doi.org/10.1145/3387906.3388625}{doi:\nolinkurl{10.1145/3387906.3388625}}


\bibitem[Robillard et~al\mbox{.}(2017)]%
        {robillard2017demand}
\bibfield{author}{\bibinfo{person}{Martin~P Robillard},
  \bibinfo{person}{Andrian Marcus}, \bibinfo{person}{Christoph Treude},
  \bibinfo{person}{Gabriele Bavota}, \bibinfo{person}{Oscar Chaparro},
  \bibinfo{person}{Neil Ernst}, \bibinfo{person}{Marco~Aur{\'e}lio Gerosa},
  \bibinfo{person}{Michael Godfrey}, \bibinfo{person}{Michele Lanza},
  \bibinfo{person}{Mario Linares-V{\'a}squez}, {et~al\mbox{.}}}
  \bibinfo{year}{2017}\natexlab{}.
\newblock \showarticletitle{On-demand developer documentation}. In
  \bibinfo{booktitle}{\emph{2017 IEEE International conference on software
  maintenance and evolution (ICSME)}}. IEEE, \bibinfo{pages}{479--483}.
\newblock


\bibitem[Roehm et~al\mbox{.}(2012)]%
        {10.5555/2337223.2337254}
\bibfield{author}{\bibinfo{person}{Tobias Roehm}, \bibinfo{person}{Rebecca
  Tiarks}, \bibinfo{person}{Rainer Koschke}, {and} \bibinfo{person}{Walid
  Maalej}.} \bibinfo{year}{2012}\natexlab{}.
\newblock \showarticletitle{How Do Professional Developers Comprehend
  Software?}. In \bibinfo{booktitle}{\emph{Proceedings of the 34th
  International Conference on Software Engineering}} (Zurich, Switzerland)
  \emph{(\bibinfo{series}{ICSE '12})}. \bibinfo{publisher}{IEEE Press},
  \bibinfo{pages}{255–265}.
\newblock
\showISBNx{9781467310673}


\bibitem[Sridhara et~al\mbox{.}(2010)]%
        {sridhara2010towards}
\bibfield{author}{\bibinfo{person}{Giriprasad Sridhara}, \bibinfo{person}{Emily
  Hill}, \bibinfo{person}{Divya Muppaneni}, \bibinfo{person}{Lori Pollock},
  {and} \bibinfo{person}{K Vijay-Shanker}.} \bibinfo{year}{2010}\natexlab{}.
\newblock \showarticletitle{Towards automatically generating summary comments
  for java methods}. In \bibinfo{booktitle}{\emph{Proceedings of the IEEE/ACM
  international conference on Automated software engineering}}.
  \bibinfo{pages}{43--52}.
\newblock


\bibitem[Sridhara et~al\mbox{.}(2011)]%
        {sridhara2011automatically}
\bibfield{author}{\bibinfo{person}{Giriprasad Sridhara}, \bibinfo{person}{Lori
  Pollock}, {and} \bibinfo{person}{K Vijay-Shanker}.}
  \bibinfo{year}{2011}\natexlab{}.
\newblock \showarticletitle{Automatically detecting and describing high level
  actions within methods}. In \bibinfo{booktitle}{\emph{2011 33rd International
  Conference on Software Engineering (ICSE)}}. IEEE, \bibinfo{pages}{101--110}.
\newblock


\bibitem[{Sun} et~al\mbox{.}(2024)]%
        {2024arXiv240707959S}
\bibfield{author}{\bibinfo{person}{Weisong {Sun}}, \bibinfo{person}{Yun
  {Miao}}, \bibinfo{person}{Yuekang {Li}}, \bibinfo{person}{Hongyu {Zhang}},
  \bibinfo{person}{Chunrong {Fang}}, \bibinfo{person}{Yi {Liu}},
  \bibinfo{person}{Gelei {Deng}}, \bibinfo{person}{Yang {Liu}}, {and}
  \bibinfo{person}{Zhenyu {Chen}}.} \bibinfo{year}{2024}\natexlab{}.
\newblock \showarticletitle{{Source Code Summarization in the Era of Large
  Language Models}}.
\newblock \bibinfo{journal}{\emph{arXiv e-prints}}, Article
  \bibinfo{articleno}{arXiv:2407.07959} (\bibinfo{date}{July}
  \bibinfo{year}{2024}), \bibinfo{numpages}{arXiv:2407.07959}~pages.
\newblock
\showeprint[arxiv]{2407.07959}~[cs.SE]
\href{https://doi.org/10.48550/arXiv.2407.07959}{doi:\nolinkurl{10.48550/arXiv.2407.07959}}


\bibitem[Sun et~al\mbox{.}(2025)]%
        {sun2025commenting}
\bibfield{author}{\bibinfo{person}{Weisong Sun}, \bibinfo{person}{Yiran Zhang},
  \bibinfo{person}{Jie Zhu}, \bibinfo{person}{Zhihui Wang},
  \bibinfo{person}{Chunrong Fang}, \bibinfo{person}{Yonglong Zhang},
  \bibinfo{person}{Yebo Feng}, \bibinfo{person}{Jiangping Huang},
  \bibinfo{person}{Xingya Wang}, \bibinfo{person}{Zhi Jin}, {et~al\mbox{.}}}
  \bibinfo{year}{2025}\natexlab{}.
\newblock \showarticletitle{Commenting Higher-level Code Unit: Full Code,
  Reduced Code, or Hierarchical Code Summarization}.
\newblock \bibinfo{journal}{\emph{arXiv preprint arXiv:2503.10737}}
  (\bibinfo{year}{2025}).
\newblock


\bibitem[Tang et~al\mbox{.}(2021)]%
        {9678882}
\bibfield{author}{\bibinfo{person}{Ze Tang}, \bibinfo{person}{Chuanyi Li},
  \bibinfo{person}{Jidong Ge}, \bibinfo{person}{Xiaoyu Shen},
  \bibinfo{person}{Zheling Zhu}, {and} \bibinfo{person}{Bin Luo}.}
  \bibinfo{year}{2021}\natexlab{}.
\newblock \showarticletitle{AST-Transformer: Encoding Abstract Syntax Trees
  Efficiently for Code Summarization}. In \bibinfo{booktitle}{\emph{2021 36th
  IEEE/ACM International Conference on Automated Software Engineering (ASE)}}.
  \bibinfo{pages}{1193--1195}.
\newblock
\href{https://doi.org/10.1109/ASE51524.2021.9678882}{doi:\nolinkurl{10.1109/ASE51524.2021.9678882}}


\bibitem[Traag et~al\mbox{.}(2019)]%
        {traag2019louvain}
\bibfield{author}{\bibinfo{person}{Vincent~A Traag}, \bibinfo{person}{Ludo
  Waltman}, {and} \bibinfo{person}{Nees~Jan Van~Eck}.}
  \bibinfo{year}{2019}\natexlab{}.
\newblock \showarticletitle{From Louvain to Leiden: guaranteeing well-connected
  communities}.
\newblock \bibinfo{journal}{\emph{Scientific reports}} \bibinfo{volume}{9},
  \bibinfo{number}{1} (\bibinfo{year}{2019}), \bibinfo{pages}{1--12}.
\newblock


\bibitem[Treude and Robillard(2016)]%
        {treude2016augmenting}
\bibfield{author}{\bibinfo{person}{Christoph Treude} {and}
  \bibinfo{person}{Martin~P Robillard}.} \bibinfo{year}{2016}\natexlab{}.
\newblock \showarticletitle{Augmenting API documentation with insights from
  stack overflow}. In \bibinfo{booktitle}{\emph{Proceedings of the 38th
  International Conference on Software Engineering}}.
  \bibinfo{pages}{392--403}.
\newblock


\bibitem[Uddin et~al\mbox{.}(2021)]%
        {uddin2021automatic}
\bibfield{author}{\bibinfo{person}{Gias Uddin}, \bibinfo{person}{Foutse Khomh},
  {and} \bibinfo{person}{Chanchal~K Roy}.} \bibinfo{year}{2021}\natexlab{}.
\newblock \showarticletitle{Automatic api usage scenario documentation from
  technical q\&a sites}.
\newblock \bibinfo{journal}{\emph{ACM Transactions on Software Engineering and
  Methodology (TOSEM)}} \bibinfo{volume}{30}, \bibinfo{number}{3}
  (\bibinfo{year}{2021}), \bibinfo{pages}{1--45}.
\newblock


\bibitem[Wan et~al\mbox{.}(2018)]%
        {wan2018improving}
\bibfield{author}{\bibinfo{person}{Yao Wan}, \bibinfo{person}{Zhou Zhao},
  \bibinfo{person}{Min Yang}, \bibinfo{person}{Guandong Xu},
  \bibinfo{person}{Haochao Ying}, \bibinfo{person}{Jian Wu}, {and}
  \bibinfo{person}{Philip~S Yu}.} \bibinfo{year}{2018}\natexlab{}.
\newblock \showarticletitle{Improving automatic source code summarization via
  deep reinforcement learning}. In \bibinfo{booktitle}{\emph{Proceedings of the
  33rd ACM/IEEE International Conference on Automated Software Engineering}}.
  \bibinfo{pages}{397--407}.
\newblock


\bibitem[Wang et~al\mbox{.}(2023a)]%
        {Wang_Li_Chen_Zhu_Lin_Cao_Liu_Liu_Sui_2023}
\bibfield{author}{\bibinfo{person}{Peiyi Wang}, \bibinfo{person}{Lei Li},
  \bibinfo{person}{Liang Chen}, \bibinfo{person}{Dawei Zhu},
  \bibinfo{person}{Binghuai Lin}, \bibinfo{person}{Yunbo Cao},
  \bibinfo{person}{Qi Liu}, \bibinfo{person}{Tianyu Liu}, {and}
  \bibinfo{person}{Zhifang Sui}.} \bibinfo{year}{2023}\natexlab{a}.
\newblock \showarticletitle{Large Language Models are not Fair Evaluators}.
\newblock  (\bibinfo{date}{May} \bibinfo{year}{2023}).
\newblock


\bibitem[Wang et~al\mbox{.}(2023b)]%
        {wang2023gdoc}
\bibfield{author}{\bibinfo{person}{Shujun Wang}, \bibinfo{person}{Yongqiang
  Tian}, {and} \bibinfo{person}{Dengcheng He}.}
  \bibinfo{year}{2023}\natexlab{b}.
\newblock \showarticletitle{gDoc: Automatic generation of structured API
  documentation}. In \bibinfo{booktitle}{\emph{Companion Proceedings of the ACM
  Web Conference 2023}}. \bibinfo{pages}{53--56}.
\newblock


\bibitem[Wei et~al\mbox{.}(2020)]%
        {10.1145/3324884.3416578}
\bibfield{author}{\bibinfo{person}{Bolin Wei}, \bibinfo{person}{Yongmin Li},
  \bibinfo{person}{Ge Li}, \bibinfo{person}{Xin Xia}, {and}
  \bibinfo{person}{Zhi Jin}.} \bibinfo{year}{2020}\natexlab{}.
\newblock \bibinfo{booktitle}{\emph{Retrieve and Refine: Exemplar-Based Neural
  Comment Generation}}.
\newblock \bibinfo{publisher}{Association for Computing Machinery},
  \bibinfo{address}{New York, NY, USA}, \bibinfo{pages}{349–360}.
\newblock
\showISBNx{9781450367684}
\urldef\tempurl%
\url{https://doi.org/10.1145/3324884.3416578}
\showURL{%
\tempurl}


\bibitem[Wong et~al\mbox{.}(2015)]%
        {wong2015clocom}
\bibfield{author}{\bibinfo{person}{Edmund Wong}, \bibinfo{person}{Taiyue Liu},
  {and} \bibinfo{person}{Lin Tan}.} \bibinfo{year}{2015}\natexlab{}.
\newblock \showarticletitle{Clocom: Mining existing source code for automatic
  comment generation}. In \bibinfo{booktitle}{\emph{2015 IEEE 22nd
  International Conference on Software Analysis, Evolution, and Reengineering
  (SANER)}}. IEEE, \bibinfo{pages}{380--389}.
\newblock


\bibitem[Woodfield et~al\mbox{.}(1981)]%
        {jss/WoodfieldSD81}
\bibfield{author}{\bibinfo{person}{Scott~N. Woodfield},
  \bibinfo{person}{Vincent~Y. Shen}, {and} \bibinfo{person}{Hubert~E.
  Dunsmore}.} \bibinfo{year}{1981}\natexlab{}.
\newblock \showarticletitle{A study of several metrics for programming effort}.
\newblock \bibinfo{journal}{\emph{J. Syst. Softw.}} \bibinfo{volume}{2},
  \bibinfo{number}{2} (\bibinfo{year}{1981}), \bibinfo{pages}{97--103}.
\newblock
\href{https://doi.org/10.1016/0164-1212(81)90029-7}{doi:\nolinkurl{10.1016/0164-1212(81)90029-7}}


\bibitem[Xia et~al\mbox{.}(2018)]%
        {icse/XiaBLXHL18}
\bibfield{author}{\bibinfo{person}{Xin Xia}, \bibinfo{person}{Lingfeng Bao},
  \bibinfo{person}{David Lo}, \bibinfo{person}{Zhenchang Xing},
  \bibinfo{person}{Ahmed~E. Hassan}, {and} \bibinfo{person}{Shanping Li}.}
  \bibinfo{year}{2018}\natexlab{}.
\newblock \showarticletitle{Measuring program comprehension: a large-scale
  field study with professionals}. In \bibinfo{booktitle}{\emph{Proceedings of
  the 40th International Conference on Software Engineering, {ICSE} 2018,
  Gothenburg, Sweden, May 27 - June 03, 2018}},
  \bibfield{editor}{\bibinfo{person}{Michel Chaudron}, \bibinfo{person}{Ivica
  Crnkovic}, \bibinfo{person}{Marsha Chechik}, {and} \bibinfo{person}{Mark
  Harman}} (Eds.). \bibinfo{publisher}{{ACM}}, \bibinfo{pages}{584}.
\newblock
\href{https://doi.org/10.1145/3180155.3182538}{doi:\nolinkurl{10.1145/3180155.3182538}}


\bibitem[Yang et~al\mbox{.}(2025)]%
        {yang2025docagent}
\bibfield{author}{\bibinfo{person}{Dayu Yang}, \bibinfo{person}{Antoine
  Simoulin}, \bibinfo{person}{Xin Qian}, \bibinfo{person}{Xiaoyi Liu},
  \bibinfo{person}{Yuwei Cao}, \bibinfo{person}{Zhaopu Teng}, {and}
  \bibinfo{person}{Grey Yang}.} \bibinfo{year}{2025}\natexlab{}.
\newblock \showarticletitle{DocAgent: A Multi-Agent System for Automated Code
  Documentation Generation}.
\newblock \bibinfo{journal}{\emph{arXiv preprint arXiv:2504.08725}}
  (\bibinfo{year}{2025}).
\newblock


\bibitem[Zhang et~al\mbox{.}(2020)]%
        {9284039}
\bibfield{author}{\bibinfo{person}{Jian Zhang}, \bibinfo{person}{Xu Wang},
  \bibinfo{person}{Hongyu Zhang}, \bibinfo{person}{Hailong Sun}, {and}
  \bibinfo{person}{Xudong Liu}.} \bibinfo{year}{2020}\natexlab{}.
\newblock \showarticletitle{Retrieval-based Neural Source Code Summarization}.
  In \bibinfo{booktitle}{\emph{2020 IEEE/ACM 42nd International Conference on
  Software Engineering (ICSE)}}. \bibinfo{pages}{1385--1397}.
\newblock


\bibitem[Zhang et~al\mbox{.}(2023)]%
        {zhang2023software}
\bibfield{author}{\bibinfo{person}{Yiran Zhang}, \bibinfo{person}{Zhengzi Xu},
  \bibinfo{person}{Chengwei Liu}, \bibinfo{person}{Hongxu Chen},
  \bibinfo{person}{Jianwen Sun}, \bibinfo{person}{Dong Qiu}, {and}
  \bibinfo{person}{Yang Liu}.} \bibinfo{year}{2023}\natexlab{}.
\newblock \showarticletitle{Software architecture recovery with information
  fusion}. In \bibinfo{booktitle}{\emph{Proceedings of the 31st ACM Joint
  European Software Engineering Conference and Symposium on the Foundations of
  Software Engineering}}. \bibinfo{pages}{1535--1547}.
\newblock


\bibitem[Zhi et~al\mbox{.}(2015)]%
        {jss/ZhiGSGSR15}
\bibfield{author}{\bibinfo{person}{Junji Zhi}, \bibinfo{person}{Vahid
  Garousi{-}Yusifoglu}, \bibinfo{person}{Bo Sun}, \bibinfo{person}{Golara
  Garousi}, \bibinfo{person}{S.~M. Shahnewaz}, {and}
  \bibinfo{person}{G{\"{u}}nther Ruhe}.} \bibinfo{year}{2015}\natexlab{}.
\newblock \showarticletitle{Cost, benefits and quality of software development
  documentation: {A} systematic mapping}.
\newblock \bibinfo{journal}{\emph{J. Syst. Softw.}}  \bibinfo{volume}{99}
  (\bibinfo{year}{2015}), \bibinfo{pages}{175--198}.
\newblock
\href{https://doi.org/10.1016/J.JSS.2014.09.042}{doi:\nolinkurl{10.1016/J.JSS.2014.09.042}}


\end{thebibliography}

\end{document}